*Journal of Physics: Photonics* Roadmap

# Quantum cascade laser roadmap

**Carlo Silvestri**[1,2,31,32]**, Aleksandar D. Rakić**[1,31]**, Dragan Indjin**[3,31]**, Ali Khalatpour**[4]**, Christian Jirauschek**[5]**, Aleksandar Demic**[3]**, Zoran Ikonic**[3]**, Paul Dean**[3]**, Nikola Vuković**[6]**, Jelena Radovanović**[6]**, Lianhe Li**[3]**, Edmund Linfield**[3]**, Michael Jaidl**[7]**, Karl Unterrainer**[7]**, Giacomo Scalari**[8]**, Jérôme Faist**[8]**, Lorenzo Luigi Columbo**[9]**, Massimo Brambilla**[10,11]**, Marco Piccardo**[12,13,14]**, Sukhdeep Dhillon**[15]**, Mithun Roy**[16]**, David Burghoff**[16]**, Karl Bertling**[1]**, Jari Torniainen**[1]**, Xiaoqiong Qi**[1]**, Thomas Taimre**[17]**, Olivier Spitz**[18]**, Frédéric Grillot**[19,20]**, Marilena Giglio**[21]**, Angelo Sampaolo**[21]**, Pietro Patimisco**[21]**, Vincenzo Spagnolo**[21]**, Eva A. A. Pogna**[22]**, Xiao Guo**[1]**, Rainer Hillenbrand**[23,24]**, Mengkun Liu**[25,26]**, Michael Brünig**[1]**, Adrian Cernescu**[27]**, Alexander A. Govyadinov**[27]**, Tecla Gabbrielli**[28,29]**, Jacopo Pelini**[28,29,30]**, Irene La Penna**[28,29]**, Alessia Sorgi**[28,29]**, Paolo De Natale**[28,29]**, Davide Mazzotti**[28,29]**, Iacopo Galli**[28,29]**, Luigi Consolino**[28,29]**, Francesco Cappelli**[28,29]**, and Simone Borri**[28,29]

[1] School of Electrical Engineering and Computer Science, The University of Queensland, Brisbane, QLD 4072, Australia
[2] Institute of Photonics and Optical Science (IPOS), School of Physics, The University of Sydney, NSW 2006, Australia
[3] School of Electronic and Electrical Engineering, University of Leeds, Woodhouse Lane, LS2 9JT, Leeds, UK
[4] Department of Applied Physics, Edward L. Ginzton Laboratory, Stanford University, Stanford, California 94305, USA
[5] TUM School of Computation, Information and Technology, Technical University of Munich (TUM), Hans-Piloty-Str. 1, 85748 Garching, Germany
[6] School of Electrical Engineering, University of Belgrade, Bulevar kralja Aleksandra 73, 11120 Belgrade, Serbia
[7] Photonics Institute, TU Wien, Gusshausstrasse 27-29, 1040 Vienna, Austria
[8] Institute for Quantum Electronics, ETH Zurich, CH-8093 Zurich, Switzerland
[9] Dipartimento di Elettronica e Telecomunicazioni, Politecnico di Torino, 10129 Torino, Italy
[10] Dipartimento di Fisica Interateneo, Università e Politecnico di Bari, 70126 Bari, Italy
[11] CNR-Istituto di Fotonica e Nanotecnologie, UOS Bari, 70126 Bari, Italy
[12] Harvard John A. Paulson School of Engineering and Applied Sciences, Harvard University, Cambridge, MA 02138, USA
[13] Department of Physics, Instituto Superior Técnico Universidade de Lisboa, 1049-001 Lisbon,
[14] Instituto de Engenharia de Sistemas e Computadores – Microsistemas e Nanotecnologias (INESC MN), 1000-029 Lisbon, Portugal[9] Dipartimento di Elettronica e Telecomunicazioni, Politecnico di Torino, 10129 Torino, Italy
[15] Laboratoire de Physique de l'Ecole Normale Supérieure, ENS, Université PSL, CNRS, Sorbonne Université, Université de Paris-Cité, 24 rue Lhomond, 75005 Paris, France
[16] Chandra Department of Electrical and Computer Engineering, Cockrell School of Engineering, The University of Texas at Austin, Austin, Texas 78712, USA
[17] School of Mathematics and Physics, The University of Queensland, Brisbane, QLD 4072, Australia
[18] CREOL, College of Optics and Photonics, University of Central Florida, Orlando, FL 32816, USA
[19] LTCI Télécom Paris, Institut Polytechnique de Paris, 19 place Marguerite Perey, Palaiseau, 91120, France
[20] Center for High Technology Materials, University of New-Mexico, 1313 Goddard SE, Albuquerque, NM 87106, USA
[21] PolySense Lab, Dipartimento Interateneo di Fisica, University and Polytechnic of Bari, Via Amendola 173, Bari 70126, Italy
[22] Istituto di Fotonica e Nanotecnologie, Consiglio Nazionale delle Ricerche (CNR-IFN), P.zza L. da Vinci 32, 20133 Milano, Italy
[23] CIC nanoGUNE BRTA and EHU/UPV, Donostia-San Sebastián, Spain
[24] Ikerbasque, Basque Foundation for Science, Bilbao, Spain
[25] Department of Physics and Astronomy, Stony Brook University, Stony Brook, NY, USA
[26] National Synchrotron Light Source II, Brookhaven National Laboratory, Upton, NY, USA

[27] Attocube systems AG (neaspec), Eglfinger Weg 2, 85540 Haar (München), Germany
[28] CNR-INO - Istituto Nazionale di Ottica, Via Carrara, 1 - 50019 Sesto Fiorentino FI, Italy
[29] LENS - European Laboratory for Non-Linear Spectroscopy, Via Carrara, 1 - 50019 Sesto Fiorentino FI, Italy
[30] University of Naples Federico II, Corso Umberto I, 40 - 80138 Napoli, Italy
[31] Guest Editors of the Roadmap.
[32] Author to whom any correspondence should be addressed.

E-mails: c.silvestri@uq.edu.au, a.rakic@uq.edu.au, D.Indjin@leeds.ac.uk

## Abstract

Quantum cascade lasers (QCLs) are unipolar semiconductor lasers first demonstrated in 1994. Since then, they have played a central role in advancing mid-infrared and terahertz photonics, becoming among the most reliable light sources in these regions of the electromagnetic spectrum. Their importance is further reinforced by their ability to generate self-starting optical frequency combs, whose investigation is motivated both by fundamental physics and by a wide range of applications, including molecular spectroscopy and free-space optical communications.

This Roadmap provides a unified overview of current advances and emerging directions in QCL research. The chapters are organized into three main sections: device design and technology; frequency combs and pulse formation; and applications of QCLs. Each chapter reviews the relevant background, summarizes the current state of the art, and identifies key challenges and future directions within its specific research area.

## Contents

# 1. Introduction

Carlo Silvestri [a,\*], Aleksandar D. Rakic [b], Dragan Indjin [c]

[a] c.silvestri@uq.edu.au, School of Electrical Engineering and Computer Science, The University of Queensland, Brisbane, QLD 4072, Australia
[b] a.rakic@uq.edu.au, School of Electrical Engineering and Computer Science, The University of Queensland, Brisbane, QLD 4072, Australia
[c] D.Indjin@leeds.ac.uk, School of Electronic and Electrical Engineering, University of Leeds, Woodhouse Lane, LS2 9JT, Leeds, UK
\*Current address: Institute of Photonics and Optical Science (IPOS), School of Physics, The University of Sydney, NSW 2006, Australia

Since their invention in 1994 [1], quantum cascade lasers (QCLs) have revolutionized the field of nonlinear photonics, becoming among the most reliable light sources in the mid-infrared (mid-IR) and terahertz (THz) [2] regions of the electromagnetic spectrum. Their uniqueness stems from their unipolar nature: their operation relies exclusively on electronic transitions within the conduction band. In QCLs, the active region consists of a heterostructure formed by a periodic sequence of quantum wells—semiconductor layers only a few nanometers thick—arranged into multiple stages. By precisely tailoring the thickness of these layers, one can engineer the energy spacing between the quantized electronic states, or subbands, confined within the wells. This ability to control subband energies enables QCLs to reach spectral regions, such as the mid-IR and THz, that are inaccessible to conventional semiconductor diode lasers [3]. This versatility has led to the development of a wide range of QCL designs, each optimized for specific applications and supported by the advancement of fabrication techniques such as molecular beam epitaxy (MBE) [4].

Another remarkable feature of QCLs, unveiled almost two decades after their invention, is their ability to generate, in a self-starting manner, optical frequency combs—broadband coherent regimes that act as frequency rulers in the spectral domain and correspond to perfectly periodic signals in the time domain [5,6]. Since their discovery in 2012, QCL-based frequency combs have exhibited distinctive characteristics, including strong frequency modulation leading to a linear chirp, accompanied by nearly constant intensity profiles in Fabry–Pérot configurations. Moreover, in ring geometries, the emission of optical solitons has been demonstrated, establishing a close analogy with microcombs generated in driven passive microresonators operating in the near-infrared range [7].

Research on QCLs is thriving, encompassing a broad spectrum of topics from device structure modelling and frequency comb dynamics to experimental studies of comb regimes and pulse generation, as well as diverse applications in imaging, spectroscopy, optical communications, and quantum light sources [8].
This Roadmap provides an integrated overview of the current landscape and emerging directions in QCL research, bridging fundamental physics, device engineering, and applications. It is organized into eighteen chapters divided into three main sections. The first section outlines the state of the art in QCL design and the latest technological advances in device fabrication. The second section explores the generation and modelling of self-starting frequency combs and mode-locked pulse regimes, including active mode-locking approaches. The third section is devoted to QCL applications, spanning THz imaging for cancer tissue detection, optical communications based on chaotic light, the characterization of quantum light states, and the implementation of near-field optical microscopy.

The convergence of theoretical, technological, and experimental efforts across the QCL community will be crucial to fully unlocking the potential of these devices, driving advances toward long-standing goals such as on-chip comb integration, miniaturized mid-IR and THz photonic circuits, and room-temperature operation. At the same time, cross-fertilization with concepts from microresonator physics and broadband photonics is expected to open new regimes of frequency-comb generation and pulse formation directly within semiconductor platforms.

## 2.1 Progress and Prospects of Terahertz Quantum Cascade Lasers Toward Room- Temperature Operation

Ali Khalatpour, Department of Applied Physics, Edward L. Ginzton Laboratory, Stanford University, Stanford, California 94305, USA

akhalat@stanford.edu

**Status**

The unique properties of materials in the 0.5–10 THz range have led to the widespread adoption of THz technology in various fields. For instance, THz radiation has been used to probe exotic quantum states—such as spin liquids, superconductors, and topological materials [1]—and in industrial and commercial applications, including quality control in manufacturing, measurement of coating thickness in the pharmaceutical industry, and defect inspection in the automotive and aerospace sectors [2], [3]. Because of the challenges in generating THz waves, there is no universal source suitable for all applications; each THz source is often tailored to specific tasks. In the frequency range of about 1.5–5 THz, terahertz quantum cascade lasers (THz QCLs) have emerged as a powerful source [4], offering high output power and adaptability for use in frequency combs and THz amplifiers [5]. Figure 1a shows the maximum operating temperature of THz QCLs ($T_{max}$) over time, revealing early rapid gains and a subsequent plateau around 200 K [6]. Initially, THz QCLs required cryogenic cooling, but progress led to operation using Stirling coolers and, more recently, single-stage thermoelectric coolers. These advancements have reduced cooling requirements and enabled more compact THz QCL systems [7], [8]. The high output power and compact form factor of Stirling-cooled THz QCL platforms [9], [10] have made them especially appealing for astrophysical applications. Their importance in astrophysics continues to grow, evidenced by NASA's ongoing investment in THz technologies. For example, the GUSTO mission—a balloon-borne observatory designed to detect THz radiation from the interstellar medium—underscores the critical role of THz QCLs in advanced astronomical research [11]. Figure 1b shows the GUSTO balloon and a 4.744 THz QCL, cooled using a Stirling cooler, highlighting these cutting-edge developments in THz technology.

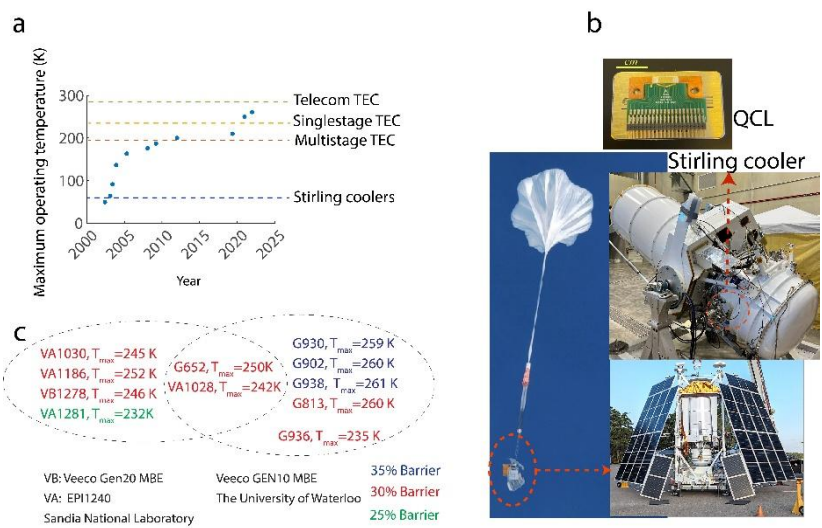

Figure1. (a) The maximum operating temperature of THz QCLs reported over the years. (b) The GUSTO balloon observatory with a 4.744 THz QCL onboard, cooled by a Stirling cooler. Credit: University of Arizona/NASA. (c) The reported $T_{max}$ identified with each unique MBE wafer label, categorized by facility, and MBE chamber [7],[8].

**Current and Future Challenges**

Temperatures above 195 K can be achieved using multistage thermoelectric coolers (TECs) [6], [12]. However, the high heat dissipation of THz QCLs significantly reduces the duty cycle—and thus the average power—when multistage coolers are used, as only cryogenically cooled detectors can be employed. Recent advances have enabled THz QCLs to operate on single-stage TECs at power levels compatible with room-temperature detectors, facilitating the development of compact, portable THz imaging systems [7]. The next major milestone for THz QCLs is likely their operation with telecommunications-grade thermoelectric coolers. Achieving this goal would greatly impact imaging and communications applications, allowing these devices to be integrated into standard butterfly packages and simplifying deployment. Techniques such as sidewall metal-coating of THz QCLs for lateral heat removal and bonding onto substrates like aluminum nitride, silicon carbide, or CVD diamond could help address thermal bottlenecks. Achieving a higher $T_{max}$ would increase average power at lower temperatures and expand the $T_{max}$ beyond the demonstrated 3.6–4.4 THz range.

Figure 1c shows the highest reported $T_{max}$ for each wafer grown by molecular beam epitaxy (MBE), categorized by facility and barrier aluminum composition (aluminum fractions of 0.25, 0.3, and 0.35 in aluminum gallium arsenide barriers). The overlap (G652 and VA1028) represents an identical design grown at both facilities. The direct phonon depopulation scheme [8] remains critical for higher-temperature operation, and the designs in Figure 1c follow this scheme to depopulate the lower lasing level. The highest $T_{max}$ has been achieved with short module lengths and a so-called two-well injector design [7], [8]. A key observation is that higher aluminum composition in barriers (i.e., taller barriers) consistently yields higher $T_{max}$, likely by reducing coupling into higher-energy parasitic channels [7], [8]. However, performance appears to have plateaued. Including additional barriers can reduce the electric field required for proper alignment of laser levels and decrease scattering rates into parasitic channels—such as in three-well designs (e.g., G936)—but has yet to improve performance [8].

**Advances in Science and Technology to Meet Challenges**

Nonequilibrium Green Function (NEGF) simulations [13] highlight the role of interface roughness, impurity scattering, and electron-electron scattering [8], all of which worsen with taller barriers and increased doping. Future efforts will likely focus on optimizing barrier thickness to improve transparent injection and reduce thermally activated LO-phonon channels, alongside systematic doping optimization. Modeling these mechanisms, particularly interface roughness, remains a significant challenge, as parameter tuning can justify nearly any result in recent studies, highlighting the complexity of this field. Another observation is that MBE quality significantly influences temperature performance [8], [14], making it difficult to distinguish whether improvements result from design, growth, or nanofabrication. However, within the same MBE chamber, better designs and fabrication processes can yield superior results. For instance, G652 outperforms G552 despite lower growth quality [7], suggesting that design can compensate for growth limitations. Further progress, if achievable, will require fundamental growth studies, for which THz QCLs provide a unique research platform. In particular, improving flux stability for aluminum in MBE systems, if feasible, may be essential for achieving room-temperature operation in addition to an "optimum" design guided by more predictive simulation tools.

**Acknowledgements**

The author would like to express gratitude to Professor Zbig Wasilewski for insightful discussions on MBE growth and to Dr. Jian-Rong Gao and Dr. Abram Gabriel Young for the GUSTO balloon images.

## 2.2 Modelling of high-performing terahertz quantum cascade lasers

Christian Jirauschek

TUM School of Computation, Information and Technology, Technical University of Munich (TUM), Hans-Piloty-Str. 1, 85748 Garching, Germany

E-mail: jirauschek@tum.de

**Status**

Modelling plays a pivotal role for the development of QCLs [1]. A central aspect is wave function engineering, i.e., tailoring the electronic and optical properties of the QCL active region by targeted engineering of the quantized electronic states in the multi-quantum-well heterostructure. In fact, the original proposal of the QCL was directly guided by this concept [2], using a density matrix (DM) formalism for describing the basic operating principle. For design development, the eigenenergies and wave functions are commonly computed with a Schrödinger-Poisson solver [1], which considers many-electron effects in Hartree approximation. Key parameters such as lasing frequency and transition dipole moment can then straightforwardly be extracted, and optimized by iteratively changing the heterostructure design.

Careful numerical optimization is crucial in particular for THz QCLs, since selective electron injection and extraction is hampered by the small energy separation between the two lasing levels. Fully quantitative design optimization requires modelling of the optical gain. This is achieved by self-consistent carrier transport simulations, which consider the relevant scattering mechanisms based on microscopic models and thus do not require empirical or fitting parameters [1]. The development of these methods is largely driven by the ongoing efforts to improve THz QCLs for practical use, including the realisation of room temperature operation. For a classification of the different approaches, see figure 1. The development of the first THz QCL was assisted by the Monte Carlo (MC) technique [3], which is a semiclassical, Boltzmann-type carrier transport modelling approach based on stochastic evaluation of the scattering transitions between the quantized states (figure 1(a)). By considering the electron wavevector **k** associated with the free in-plane motion, MC fully accounts for intrasubband scattering and computes the **k** dependent electron distributions [1]. However, coherent carrier transport is not properly considered. Subsequently, a hybrid DM-MC scheme was developed to include resonant tunnelling into MC at least in rudimentary form [4] (figure 1(b)). Coherent and scattering transport are both fully considered in quantum transport methods (figure 1(c)) such as the nonequilibrium Green's functions (NEGF) approach [1,5,6], which has been the method of choice for the development of recent THz QCL record temperature designs [6,7]. NEGF is the most general, but also the most computationally expensive of these approaches. Consequently, various **k**-resolved DM-based transport models have been introduced in an effort to balance rigor and numerical load [8-10].

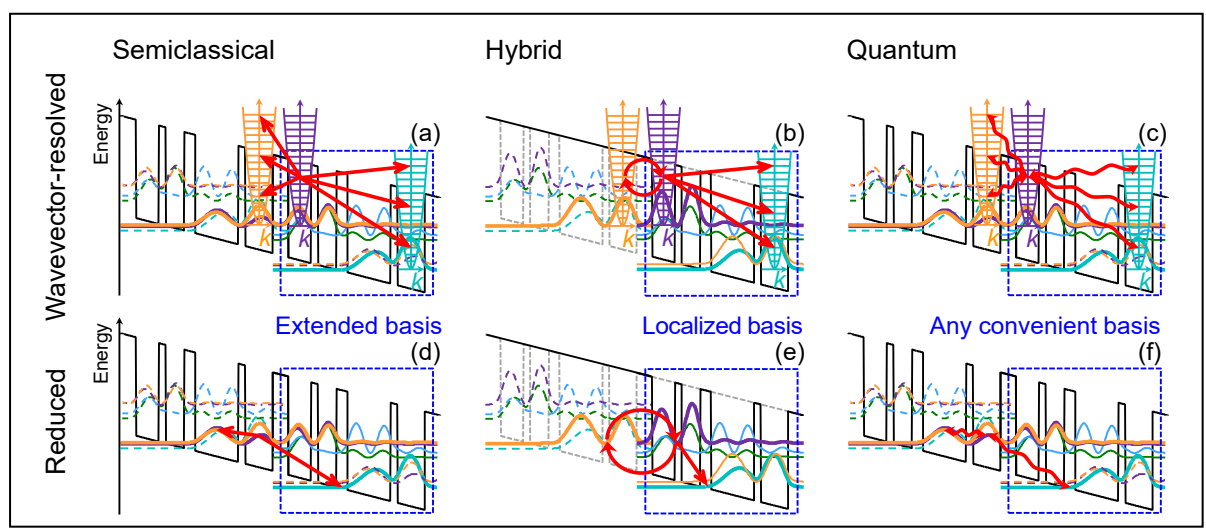

**Figure 1.** Classification of carrier transport models based on an exemplary THz QCL structure (for some states the energy dispersion relation is illustratively sketched). Wavevector-resolved models use electronic eigenstates characterized by the corresponding wave function and in-plane wavevector **k**. For reduced approaches the **k** dependence is removed. Semiclassical methods describe transport between the quantized states by incoherent scattering (straight arrows), while hybrid approaches also include resonant tunnelling through thick barriers (circular arrows). Quantum transport models fully consider coherences between the states (wavy arrows).

**Current and Future Challenges**

For the existing carrier transport modelling approaches, there is always a trade-off between accuracy and versatility on one hand, and numerical efficiency on the other hand. For optimization tasks, semiclassical rate equation models are still widely used [11,12]. These are much faster than MC since the **k** dependence has been removed by suitable averaging of the scattering rates (figure 1(d)) [1,11]. Although accuracy can be improved by employing **k**-averaged hybrid (figure 1(e)) [13] and DM (figure 1(f)) [14] approaches instead, such reduced models are inherently limited since they do not account for intrasubband processes. Another strategy is to neglect computationally expensive carrier transport mechanisms. Besides omitting coherence effects in semiclassical models, also quantum transport approaches typically rely on approximations to make computations feasible: In particular, electron-electron scattering is often only treated in Hartree approximation, since its implementation as a two-body interaction considerably complicates computation [1,5,6,8-10,13,14]. A further strategy for reducing the numerical load is to use an adapted basis set which is computationally favourable for a certain design. For example, localized wave functions based on artificially subdividing the heterostructure in scattering transport regions, separated by tunnelling barriers, have been employed in hybrid DM-MC [4] and reduced [13] models (figure 1(b),(e)). Especially for THz QCLs, the identification of tunnelling barriers can be ambiguous [10]. By using so-called EZ states, this problem is avoided at the cost of introducing an empirical threshold energy [15]. A general issue associated with any finite basis set is that the simulation result depends on the chosen basis since the completeness relation is not entirely fulfilled, impeding numerical comparison of different designs [1,5,10]. This issue has for example been avoided in NEGF by using a spatial grid for discretization rather than adapted basis functions, at the expense of a higher numerical load [1]. Besides the development of increasingly powerful simulation approaches, improved numerical QCL design and optimization will greatly rely on the growth of numerical resources and the use of refined optimization strategies [6,12,16].

To date, carrier transport models have mainly focused on calculating the unsaturated gain, while only few works have dealt with simulating the actual laser operation [1,5,13]. Such approaches are however required for tasks such as maximizing the output power and wall-plug efficiency of THz QCLs. These aspects will gain importance once further progress has been made regarding the temperature performance. Also, electrothermal modelling will become relevant especially for continuous-wave operation [17].

**Advances in Science and Technology to Meet Challenges**

A standard numerical optimization task consists in maximizing a merit function, for example the peak gain, for a specific (e.g., two- or three-well) THz QCL design, by varying certain parameters such as layer thicknesses [6]. Additionally, a bias sweep is required for each point in parameter space since the merit function also depends on the applied voltage. For enhanced predictive power, the use of elaborate, numerically expensive modelling techniques such as NEGF is advisable. To reduce the numerical load, this will require the targeted application of suitable optimization schemes, rather than performing brute force parameter sweeps [6,12,16]. A recent study indicates that Bayesian optimization is especially efficient in this context, and can also evaluate the robustness of a design towards model and growth inaccuracies [6]. Furthermore, the application of machine learning techniques for design purposes shows great promise [12]. Apart from refining optimization strategies, design optimization will greatly benefit from the growth of numerical resources. In this context, unlocking the full potential of graphics processing units (GPUs) for carrier transport calculations bears enormous potential [16]. Increased computational power will also allow for the development of even more accurate and versatile simulation approaches. For example, the models are to date usually restricted to a single or few QCL stages, assuming strict periodicity. This assumption is violated if domain formation occurs, which tends to negatively affect QCL performance, necessitating the simulation of multiple stacks. A suitable approach with linear scaling between computational complexity and stack number has for example been demonstrated based on the DM method [9]. Likewise, the growth of computational power will enable sufficiently accurate spatial discretization especially in quantum transport approaches, avoiding the need of restrictive basis sets and allowing adequate consideration of carrier leakage into continuum states, which especially affects high temperature operation and poses challenges in modelling [7,15]. In the longer run, the development of quantum simulation platforms may offer an alternative route for QCL design simulation and optimization [18].

With improved models and optimization strategies as well as increasingly powerful computational resources, refined device optimization becomes possible by increasing the parameter space, e.g., varying barrier heights and doping profiles in addition to layer widths. Furthermore, advanced simulation techniques without design-specific basis sets also allow for quantitative comparisons between different designs [6]. Thus, as an ultimate goal, simulations may be used for an automated exploration of unconventional QCL designs, potentially employing alternative material systems [11].

**Concluding Remarks**

Self-consistent carrier transport simulations only require the design specifications and well-known material parameters as input, and are thus the method of choice for quantitative simulations of THz QCLs. Current methodological development largely focuses on a realistic incorporation of both quantum and scattering transport while keeping the numerical load manageable. Since the numerical effort typically increases with accuracy and versatility of the simulation approach, the exploitation of advanced numerical optimization methods and computational resources will play a key role for improved numerical design. Besides optimization of the stationary performance, another important aspect is THz QCL development for mode-locked short-pulse or comb operation. Proper modelling of mode-locking requires spatiotemporally resolved simulations over many cavity roundtrips. For this purpose, numerically efficient semi-phenomenological models such as the Maxwell-Bloch equations are typically used [19]. A quantitative approach can be obtained by coupling a dynamic Maxwell-DM model, based on a generalized multilevel Hamiltonian, to carrier transport simulations at the operating point, which provide the scattering and dephasing rates as input for the dynamic model [20]. With the further growth of numerical resources, also the direct application of advanced carrier transport models to long-term dynamic QCL simulations will become realistic [5,8,9,10,14].

**Acknowledgements**
This work was supported by the European Union's QuantERA II (G.A. n. 101017733) – QATACOMB project "Quantum correlations in terahertz QCL combs" (funding organisation: Deutsche Forschungsgemeinschaft – Germany under project number 491801597).

## 2.3 Design of terahertz quantum cascade lasers

Aleksandar Demic, Zoran Ikonic, Paul Dean and Dragan Indjin
School of Electronic and Electrical Engineering, University of Leeds, Woodhouse Lane, LS2 9JT, Leeds, UK
E-mail: A.Demic@leeds.ac.uk

**Status**

THz QCLs lase across the 1.2–5.7 THz range [1], which corresponds to an energy range of 5.8–23.6 meV. They can employ up to 20 intersubband levels within a single period of their superlattice. These energy levels are best envisioned as discrete two-dimensional, "nearly parabolic baskets" of states that electrons can occupy, which form due to the crystalline nature of the semiconductor heterostructure. The concept of effective lasing schemes [2], commonly used to illustrate the lasing principle in any active medium, provides fundamental insight into the key traits and classifications of THz QCLs [3]. We can formulate the effective lasing schemes for common THz QCL designs (Fig. 1) by representing the intersubband minima as discrete energy levels and assuming that any cluster of narrowly spaced levels can be effectively treated as a single energy level.

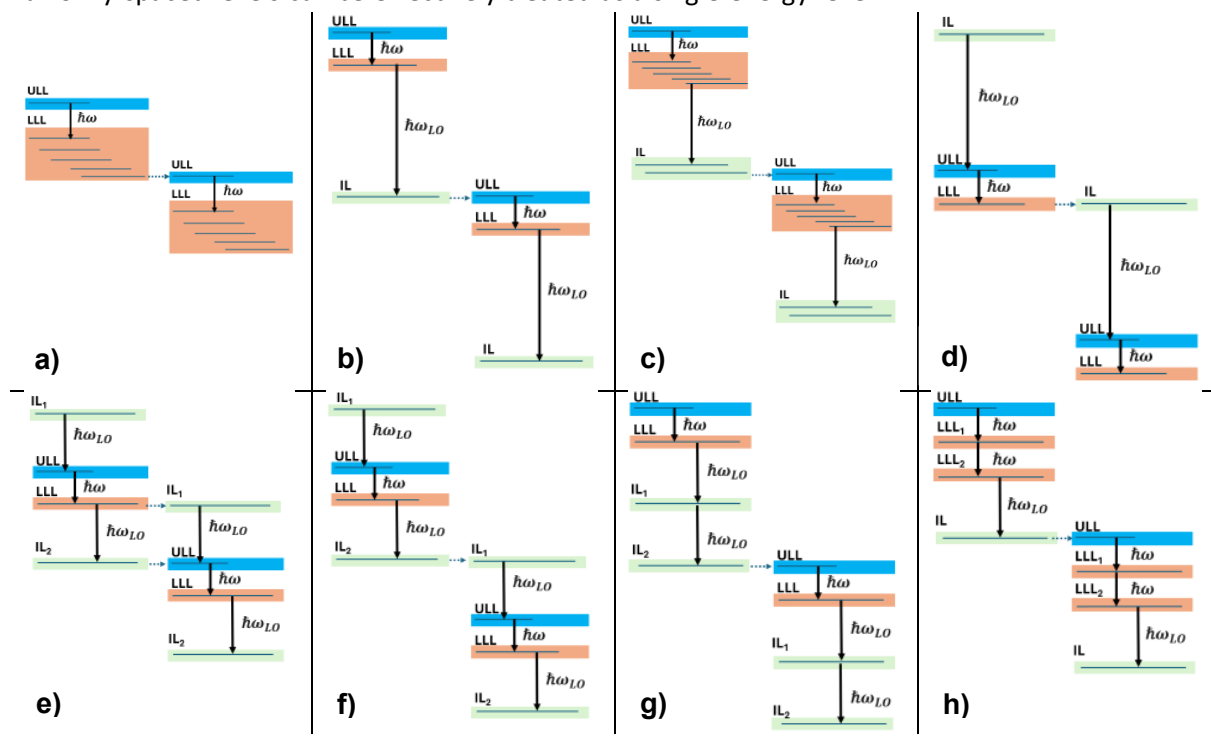

**Figure 1**. a) Bound-to-continuum (BTC) design, b) Resonant-phonon design (RP), c) Hybrid design, d) Scattering assisted design, e) Dual-resonance phonon-photon-phonon design, f) Phonon-photon-phonon design, g) Double phonon design, h) Dual lasing design. Two THz QCL periods are presented for each design, along with their effective upper lasing level (ULL), lower lasing level (LLL) and injection level (IL).

The BTC, RP, and Hybrid designs (Fig. 1) comprise the majority of experimentally realized THz QCLs, whereas the others [3] have been demonstrated mainly as proofs of concept. Both resonant tunnelling and longitudinal-optical (LO) phonon scattering mechanisms can be used to efficiently pump or extract the lasing states. The resonant electric field bias, $K_R$ (proportional to the applied voltage), satisfies the tunnelling condition $K_R = \frac{1}{L}\frac{\Delta E}{e}$ where $L$ is the superlattice period length and $\Delta E = E_{ULL} - E'_{IL}$ is the energy difference between the tunnelling states. Designs that utilize LO-phonon scattering mechanism typically have large $\Delta E$ and low $L$ and may exhibit large threshold whereas designs that employ a larger number of quantum wells (low $\Delta E$ and high $L$) and may have lower thresholds. LO-phonon-based designs are generally more robust in terms of temperature performance, while designs with long $L$ tend to form minibands due to narrowly spaced level clusters, which typically degrade temperature performance. For this reason, the BTC structure typically has low thresholds, whereas RP designs exhibit high thresholds. However, the presence of minibands in BTC designs limits continuous-wave (CW) operation to around 50 K [4], while RP designs, although rarely operating in CW, have been demonstrated up to 261 K [5] in pulsed mode. Hybrid designs offer a compromise by compensating for a larger $\Delta E$ with a long superlattice period $L$, resulting in moderate thresholds and higher CW operating temperatures (up to 129 K) [6]. These characteristics make Hybrid designs the most frequently used structures in THz QCL applications.

**Current and Future Challenges**

While there are numerous THz QCL designs [7], the key challenge remains improving their temperature performance and optical power at elevated temperatures in both pulsed and CW operation. Pulsed operation has nearly achieved room-temperature performance [5]; however, this comes at the cost of a very high threshold current and voltage (10 A and 40 V), due to the short superlattice period in usually used two-well designs. In contrast, the record for CW operation is 129 K [6], but the output power of CW devices decreases significantly with increasing temperature. Systematic design optimization is essential to realize Peltier-cooled (>200 K), low-threshold devices for pulsed operation and higher-power devices for CW operation.

The design of THz QCLs is heavily guided by theoretical modeling [8] which necessitates use of advanced quantum transport models such as the density matrix (DM) [9] and non-equilibrium Green's function (NEGF) [10]. NEGF provides deeper physical insight of electron transport, but its high computational cost often requiring hours per simulation makes it impractical for large-scale design exploration. In contrast, the DM model is significantly more efficient, making it better suited for targeted optimisation studies.

THz QCL performance is highly sensitive to quantum well and barrier thicknesses, with each layer introducing an additional degree of freedom in design space. For example, varying thicknesses of a 6-layer three well RP design in only 10 steps per layer results in $10^6$ combinations. As a result, DM-based simulations, especially when parallelized on high-performance computing (HPC) systems are preferred for exploring this vast parameter space. However, even with full parallelization, exhaustive search is impractical. Extending brute-force approach to designs capable of CW performance that may employ 15-20 layers results in parameter space that is computationally prohibitive.

The key future challenge lies in reducing the design space through better physical insight or optimization techniques. Promising strategies include focusing on the most critical quantum wells, decoupling regions to reduce dimensionality, or using simplified models (e.g., solving only the Schrödinger equation) to pre-select configurations that satisfy specific level schemes.

An additional future challenge involves adapting QCL designs to alternative material systems, such as the materials with larger electron-LO phonon resonant energies, for example ZnO/MgZnO [11], where new scattering mechanisms and more complex band structures may require different modelling approaches beyond the traditional 1D Schrödinger equation under the effective mass approximation.

**Advances in Science and Technology to Meet Challenges**

The most promising approach for tackling THz QCL optimization is the use of Bayesian optimization algorithms [12], which have demonstrated success in similar high-dimensional design tasks. Genetic algorithms [13] can also be considered, although they often converge to local rather than global optima. Additionally, machine learning techniques [14] may help reduce the parameter space by up to 30% and uncover correlations between active region designs as the number of layers increases.

A significant recent development involves a paradigm shift in the design of THz QCLs that utilize LO-phonon scattering mechanisms [3, 15]. Figure 2 illustrates the well-known dependence of LO-phonon scattering rates on energy level separation in GaAs-based quantum wells [15]. In the GaAs/AlGaAs material system, LO-phonon emission is resonant at approximately 36 meV—historically the preferred energy separation in most RP and Hybrid designs for depopulating the LLL. However, recent studies [3, 15] suggest that this choice may be suboptimal for high-temperature performance.

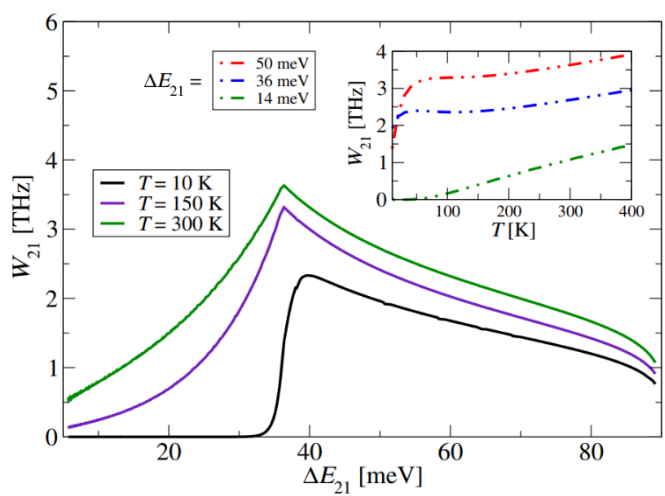

**Figure 2**. Averaged electron–LO-phonon scattering rate between first two quantum states of a single quantum well with no external bias. Energy was varied by varying well thickness. Inset illustrates the scattering rate at 14, 36 and 50 meV which correlates to level separations in 3 level RP THz QCL design.

While LO-phonon scattering is strongest at 36 meV, it remains appreciable up to 80 meV, with scattering efficiency increasing at elevated temperatures. For example, a typical three-level RP design operating at ~3.4 THz often features a 2→1 transition at 36 meV and a 3→2 transition at 14 meV. In this scenario, the upper laser level (ULL) can also scatter via a 3→1 transition at 50 meV, potentially degrading population inversion. Additionally, thermal backfilling—the repopulation of the LLL via LO-phonon absorption—becomes more likely at higher temperatures, further limiting performance if the level separation is 36 meV. The study in [3, 15] proposed that high-temperature operation could be enhanced by engineering LO-phonon transitions around 48 meV or higher, as both thermal backfilling and ULL depopulation would be reduced at elevated temperatures. The current record for pulsed operation up to 261 K [5] was achieved using LO-phonon transitions between 48–60 meV for efficient depopulation of the LLL. This represents a critical shift in design strategy and opens new possibilities for improving the temperature performance of both RP and Hybrid THz QCLs where systematic HPC-driven optimization will play a key role.

**Concluding Remarks**

The design of THz QCL devices leverages quantum mechanical tunneling and LO-phonon scattering mechanisms to efficiently maintain population inversion. Each design is highly sensitive to layer thickness variations in the heterostructure, and every additional layer introduces a new degree of freedom in the design space. Devices of practical interest—capable of continuous-wave (CW) operation with high output power—typically require a large number of layers (15–20), which poses a significant computational challenge, even when parallelizing carrier transport modeling.

A recent paradigm shift toward utilizing >36 meV depopulation transitions via electron-LO-phonon scattering has enabled operation within the Peltier-cooled regime for THz QCL applications, with lasing demonstrated up to 261 K in pulsed mode. However, these devices exhibit very high threshold currents, creating impractical electronic demands for portable THz systems. Furthermore, novel material systems [1] have recently gained attention [11], offering the potential to overcome the limitations of GaAs/AlGaAs. Systematic design approaches that can optimize the performance of THz

QCLs will be critical for advancing the field. These efforts aim to realize high-power CW THz QCLs, portable Peltier-cooled devices with moderate thresholds, and devices based on novel material systems.

**Acknowledgements**

This work was supported , European Cooperation in Science and Technology (COST) Action CA21159 PhoBioS, the Engineering and Physical Sciences Research Council (EPSRC) UK (Grants No. EP/W028921/1, No. EP/T034246/1, and No. EP/V004743/1). Numerical analysis in this work was undertaken on ARC4, part of the High Performance Computing facilities at the University of Leeds, UK.

## 2.4 Nonlinear effects in QCLs

Nikola Vuković, Jelena Radovanović,
School of Electrical Engineering, University of Belgrade, Bulevar kralja Aleksandra 73, 11120 Belgrade, Serbia
nvukovic@etf.bg.ac.rs

**Status**

*[This section provides a brief history and status, why the field is still important, what will be gained with further advances. (400 words max)]*

After three decades of rapid advancement, quantum cascade laser (QCL) has become the most important coherent light source in the mid-infrared (mid-IR, λ ≈ 3–30 μm) and terahertz (THz, λ ≈ 30–300) ranges with applications in gas sensing, spectroscopies, medical imaging, and free-space communications [1], [2]. QCL possesses strong nonlinearities due to the strong coupling between the quantum subbands, making it an ideal platform for nonlinear photonics, such as THz difference frequency generation (DFG) and direct frequency comb (FC) generation via four-wave mixing [3]. THz QCLs based on intra-cavity DFG are the only electrically pumped monolithic semiconductor light sources covering 1–6 THz at room temperature [4]. Since their first demonstration by Belkin et al. in 2007 [5] the power output of THz DFG-QCLs has increased significantly, relying on the dual-upper-state (DAU) active regions with the giant second-order nonlinear susceptibility and the Čerenkov phase-matching scheme [6]. THz DFG-QCLs use a mid-IR QCL active region specially engineered for an efficient intra-cavity non linear mixing process and produce two mid-IR pump frequencies as well as the THz frequency corresponding to the difference of the mid-IR pump frequencies [4]. Current state-of-the-art results, demonstrated by Lu et al. from Northwestern University, are room-temperature (RT) 3.5 THz DFG-QCLs with THz peak power output of 1.9 mW (pulsed mode) and the THz wall plug efficiency (WPE) of $0.7 \times 10^{-5}$ as well as RT continuous wave (CW) devices with over 14 μW of THz power output at 3.4 THz [7]. Devices employ a buried heterostructure with epi-down mounting on a diamond heat sink for reduced thermal resistance [8]. Čerenkov THz DFG-QCLs provide broadly tunable THz output that spans a range of nearly 1–6 THz employing the external cavity (EC) setup [9] or using various monolithic tuner configurations [10]. Čerenkov THz DFG-QCLs can be operated as RT FC sources with the potential of spanning the entire 1–6 THz spectral range and beyond [2]. The only room-temperature THz DFG-QCL FC at 3.0 THz was reported in [11], with a largely detuned distributed-feedback (DFB) grating integrated into the QCL cavity to select a single-mode QCL pump wavelength separated from that of a FC as shown schematically in Fig. 1a. The high-order harmonic mid-IR comb emission is down-converted by a single-line mid-IR pump into the THz frequency comb as shown in Fig. 1(b). Further development of THz DFG-QCLs and FCs is necessary for various emerging QCL-based applications in spectroscopy, sensing, and communication.

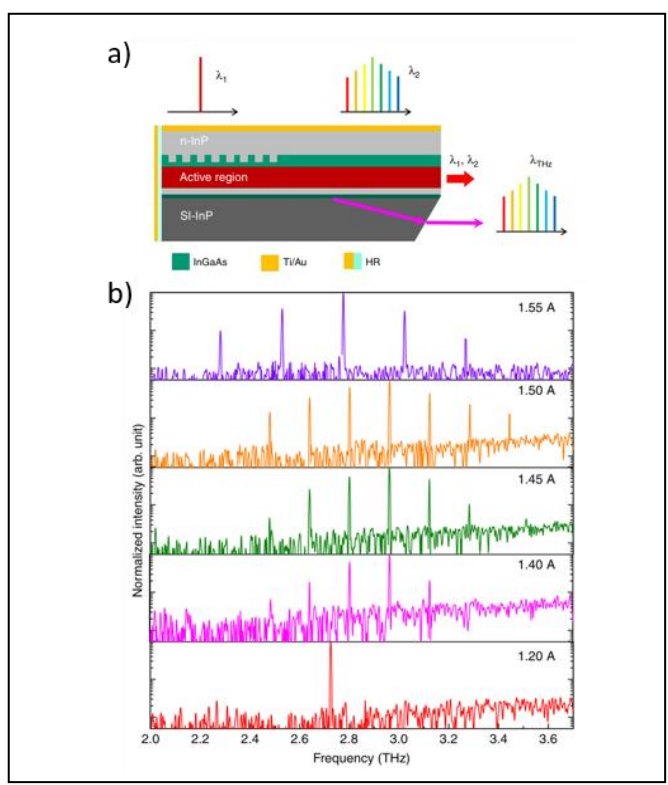

**Figure 1.** (a) Schematic of the RT THz frequency comb generation in a Čerenkov THz DFG-QCL device from [11]. Lasing THz spectra of the QCL comb from (a) evolving at different currents from 1.20 to 1.55 A at RT CW operation. Reproduced from [11] under Creative Commons Attribution International License (https://creativecommons.org/licenses/by/4.0/ ).

**Current and Future Challenges**

*[This section discusses the big research issues and challenges. (400 words max)]*

The major challenges facing THz DFG-QCLs and frequency combs are providing higher output power beyond mW for THz DFG QCL and high-precision application using the QCL frequency comb, both of which require WPE of mid-IR QCL to be elevated up to 50% [3]. Currently WPE of THz DFG QCL is only ~$0.7 \times 10^{-5}$ at room temperature and $2.7 \times 10^{-5}$ at 78 K [4]. The mid-IR power upscaling with high beam quality and single-mode operation, while maintaining nonlinear coefficient and coherence length optimal, represents the next milestone. The increase in WPE is important for reducing excessive heat and therefore increasing CW power, but achieving significantly higher output power can also simply be achieved by increasing the volume of the laser active region if the same WPE is maintained [3]. Power scaling of QCLs in CW mode is much more demanding than for pulsed mode due to the significant heat in CW. Furthermore, QCL requires some minimal voltage to align the cascades, posing a challenge to improving WPE [3]. Additionally, a lower quantum efficiency and stronger carrier thermalization during electron transport in minibands prevents the increase of QCL WPE. Achieving higher CW WPE therefore represents a trade-off between the active region confinement factor and the thermal resistivity. Semi-insulating InP substrate used in the state-of-the-art devices has greater losses compared to the semi-insulating Si substrate. Namely, in semi-insulating InP substrate THz outcoupling efficiency is below 10% of THz power generated in their active region. Therefore, direct integration of a THz DFG-QCL on a semi-insulating Si substrate would be an interesting approach for THz power enhancement. A more straightforward solution is the direct epitaxial growth of QCLs on the desired photonics platform like Si [3].

Since less than 5% of the generated THz power in the state of the art devices is outcoupled to free space, there is plenty of room for improvement in terms of optimization of the THz outcoupling efficiency. Fundamental limit on WPE for THz DFG systems can be estimated by Manley–Rowe relations [12], and knowing that the current values for room-temperature wall-plug efficiency of mid-IR QCLs are in the range of 10–20% at λ≈ 5–10 μm, one obtains the limit for WPE of 1 % for 4 THz generation in THz DFG-QCLs at RT [2].

**Advances in Science and Technology to Meet Challenges**
*[This section discusses the advances in science technology needed to address the challenges. (400 words max)]*

Improvements in terms of the THz output power and WPE are possible by optimization of the active region and waveguide designs. Namely, employment of DAU active region designs together with III–V-on-silicon hybrid laser concept is expected to produce THz peak output power of about 100 mW and increase the WPE by at least two orders of magnitude [4]. Furthermore, CW power scaling of mid-IR QCLs was successfully demonstrated in Razeghi's group at Northwestern University by using an array of narrow ridge optical phased-locked array (OPA) QCLs [13], [14]. These important results have shown that in pulsed mode, the mid-IR can be upscaled to 50 W by using a 16-element OPA design, see Fig. 2a, promising that, after mid-IR-to-THz conversion, THz power of over 100 mW could be expected in the future, see Fig. 2b, which would be enough power for many novel applications [3].

The proof-of-concept for the realization of longwave IR Si-based QCL capable of RT operation and high-power output was also demonstrated by the same group by growing InP-on-Si templates by gas source molecular-beam epitaxy (GSMBE) [15], opening possibilities for realization of multispectral (from visible to THz) single Si-based photonic integrated circuit in the future [3]

Although the Čerenkov scheme in THz DFG-QCLs resulted in a current state of the art devices, both in pulsed and CW mode, new approaches are needed for the improvement of WPE at room temperature. Limitations of the semi-insulating InP substrate extraction efficiency of THz DFG-QCLs could be overcome using double-metal THz waveguides and gratings to outcouple THz radiation from the laser waveguide, instead of a Čerenkov phase-matching scheme [16]. Nevertheless, the development of these grating-based devices is still at its initial stages, and they cannot provide broadband tuning and broadband FC generation [2].

THz DFG-QCL large-scale production and commercialization will be possible in the future since metal organic chemical vapor deposition (MOCVD) technology, famous for its high-volume and low-cost of production, was recently employed for growing high-performance QCLs [3], [17].

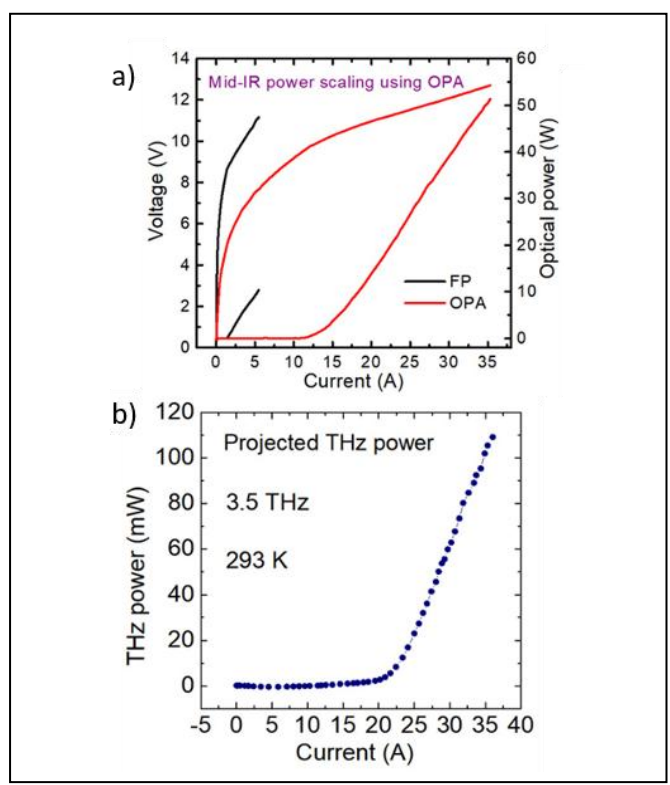

Figure 2. P-I-V and WPE characterization of a 16-element QCL OPA as functions of the current in a continuous wave operation. (a) P-I-V characterizations of a 16-element QCL OPA and a FP reference device as functions of the current in a pulsed mode. (b) Projected THz power of the THz DFG-QCL using OPA design. Reproduced from [3] under Creative Commons Attribution (CC BY) license ( https://creativecommons.org/licenses/by/4.0/ ).

**Concluding Remarks**

*[Include brief concluding remarks. This should not be longer than a short paragraph. (200 words max)]*
In this roadmap we have reviewed state of the art devices and future challenges in the field of THz DFG-QCLs and frequency combs generated by strong nonlinearities in mid-IR QCL active region, currently the only technology capable of producing a monolithic semiconductor laser source of coherent THz output operable at and above room temperature. Although THz DFG-QCLs have experienced enormous progress in terms of power output and the mid-IR-to-THz nonlinear conversion efficiency at room temperature in the past two decades, significant improvements of the existing and development of new approaches are necessary for several important applications.

**Acknowledgements**

*[Please include any acknowledgements and funding information as appropriate.]*
This work was financially supported by the Science Fund of the Republic of Serbia, Grant No. 10504, "Ultra-short pulsations from TERAhertz quantum cascade laser using passive mode-LOCKing with graphene saturable absorber"-TERALOCK, European Cooperation in Science and Technology (COST) Action CA21159 PhoBioS, and Ministry of Science, Technological Development and Innovation of the Republic of Serbia under contract number: 451-03-137/2025-03/200103.

**References**

*[(Separate from the two-page limit) Maximum 20 References. Please provide the full author list, and article title, for each reference to maintain style consistency in the combined roadmap article. Style should be consistent with all other contributions- use IEEE style]*

## 2.5 MBE of high-power GaAs/AlGaAs THz QCL structures

Lianhe Li, Edmund Linfield, School of Electronic and Electrical Engineering, University of Leeds, Woodhouse Lane, LS2 9JT, Leeds, UK

L.H.Li@Leeds.ac.uk; E.H.Linfield@leeds.ac.uk

**Status**

The performance of Terahertz (THz) quantum cascade lasers (QCLs) has advanced significantly, with peak output powers now reaching the multi-watt level [1, 2]. This progress, which builds upon foundational high-performance active region (AR) designs [3], has been driven largely by improvements in material growth by molecular beam epitaxy (MBE) [4]. A key early benchmark was 248 mW at a cryogenic temperature [5]. A subsequent breakthrough was achieved in 2013 with 470 mW using wafer bonding —a technique that combines two ARs to double the gain medium thickness [6]. While effective, this method is limited to symmetric AR designs. An alternative approach, scaling up device dimensions, first broke the 1 W barrier [1]. These inspired efforts to grow commensurately thick ARs monolithically, culminating in the demonstration of peak powers up to 2.4 W from a single, ~24 µm-thick AR grown entirely by MBE [2]. In continuous wave (CW) mode, where thermal management is critical, output power has also progressed steadily, recently reaching 312 mW [7]. These advancements since 2014, often realized with only marginal modifications to the original AR design [3], underscore the critical role of epitaxial growth quality in advancing performance limits.

**Current and Future Challenges**

The performance of THz QCLs is heavily dependent on the epitaxial realization of the AR design. Indeed, recent progress in high-power devices is primarily attributable to MBE growth improvements [4], highlighting several fundamental challenges in material growth.

First, period uniformity across the entire AR is of ultimate import. A THz QCL AR consists of hundreds of periods, each containing precisely engineered quantum wells and barriers. Any deviation from the designed layer thicknesses and composition compromises the electronic subband alignment, affecting injection efficiency and population inversion. For typical ARs approximately 10 µm thick, this demands precise control of growth rates for extended periods (12–20 hours) [4, 8]. However, in MBE effusion cells—pre-loaded with a fixed, non-replenishable charge— the source material depletes during long runs, making it difficult to maintain the required layer thickness and composition across the thousands of individual layers that form the AR. The cumulative effect of even minor drift can significantly impact device performance.

Second, material purity plays a key role. Unintentional background impurities act as sources of scattering and non-radiative recombination centers, and can create parasitic conduction channels, all of which reduce the lifetime of laser levels and overall efficiency [4]. High-performance THz QCLs require an exceptionally low background doping level in the GaAs that constitutes over 95% of the AR. The best results are achieved with p-type background doping in the low $10^{13}$ cm$^{-3}$ range [2, 4]. While MBE can achieve this, it remains a significant technical challenge requiring meticulous system preparation. In contrast, metalorganic chemical vapour deposition (MOCVD) inherently has a more complex chemical environment, typically yielding higher background doping levels ($10^{15}$–$10^{16}$ cm$^{-3}$) [9]. This is likely one of the factors in the limited success of producing high-performance THz QCLs with this technique.

Third, to enable higher operating temperatures, there is increasing interest in employing higher Al-composition barriers in AR designs to suppress carrier leakage via parasitic transport to higher-lying continuum states [10]. This trend introduces the challenge of controlling interface quality. The surface mobility of Al adatoms is lower than that of Ga adatoms, meaning that surface and interface roughness tend to increase as the Al composition in AlGaAs increases. For designs using ultra-thin pure AlAs barriers (some less than two monolayers thick) [11], there is a heightened risk of islanding, where the layer does not form a continuous film. Such roughness scatters carriers, broadening the gain spectrum and lowering peak gain. This will be a key consideration for the development of future high-performance devices.

**Advances in Science and Technology to Meet Challenges**

To address the challenges of structural precision, material purity, and interface quality, several advanced growth and characterization techniques have been refined and/or developed.

For structural precision, traditional pre-growth calibration using high-resolution x-ray diffraction (XRD) on GaAs/AlGaAs superlattices can be unreliable [12]. This is due to flux transients during shutter operations and ambiguity in determining individual layer thicknesses from the diffraction pattern. A more robust method employs thick host layers interspersed with thin marker layers [4, 13]. In a GaAs/InAs superlattice, for example, the GaAs host layers are thick (tens of nanometres) while the InAs marker is extremely thin (a single monolayer). The negligible marker growth time minimizes flux transient effects, while strain contrast yields a strong XRD signal with sharp satellite peaks. The angular separation of these peaks allows for a highly accurate and reliable determination of the period thickness, and thus the growth rate of the host material.

For real-time monitoring during the extended growth of a QCL, pyrometric spectrometry is a highly effective technique [4]. It works by measuring thermal radiation from the substrate. As layers of different materials (e.g., GaAs and AlAs) are deposited, the surface emissivity and optical interference conditions change, causing the pyrometer signal to oscillate. The period of these oscillations corresponds directly to the time taken to grow a specific layer thickness (e.g., a quarter-wavelength), providing a direct, in-situ measurement of the growth rate. Unlike RHEED oscillations, which decay after a few layers, pyrometry tracks the full growth, enabling in-situ calibration and growth rate drift compensation, allowing for precise thickness control to within ±0.5% even for a 24 μm-thick AR grown over 28 hours [2].

Achieving ultra-low background doping requires rigorous procedures to prepare the MBE system [14]. An essential step is extensive system baking (typically at ~200°C for several days) and cell outgassing after any exposure to air to remove water vapour and other contaminants. The gallium (Ga) source requires particular attention, as exposure to air forms a stubborn surface oxide ($Ga_2O_3$) that is difficult to remove. At high temperatures, this oxide layer is slowly reduced through reaction with the underlying molten Ga, which releases volatile $Ga_2O$ [15]. Until the $Ga_2O_3$ is fully consumed, this process continuously releases $Ga_2O$, which acts as a primary source of contamination. An effective strategy is to use modern two-zone Ga cells, where the tip zone is heated to a higher temperature for an extended period after material charging. This outgassing serves to substantially remove the $Ga_2O_3$ before growth commences, enabling the rapid achievement of high-purity material.

Finally, to minimize interface roughness associated with high Al-composition barriers, the optimization of growth parameters is essential [14]. High-quality AlGaAs (30-40% Al) typically requires a growth temperature >725°C, significantly higher than the ~600°C used for standard THz QCLs. This higher temperature, however, introduces risks such as Ga re-evaporation, dopant diffusion, and interface intermixing. Alternatively, a lower growth rate can be used, but this tends to increase impurity incorporation from the background and extend already long growth times. Other techniques, such as growth interruptions at interfaces to promote surface smoothing, are also options. Determining the optimal growth conditions by balancing these competing factors is a key area of ongoing research.

**Concluding Remarks**

The development of high-performance THz QCLs is driven by advancements in AR designs and is heavily dependent on their precise epitaxial realization in the mature GaAs/AlGaAs material system via MBE. While future long-term advancements may rely on exploring alternative material systems to overcome the intrinsic limitations of GaAs, a pressing challenge is the performance discrepancy among devices with nominally identical ARs grown in different laboratories [16]. Sb-based materials offer high gain [17]. InP-based QCLs have demonstrated excellent performance in the mid-infrared and are being explored for THz frequencies [18], and wide-bandgap materials like GaN offer the potential for high-temperature operation due to their large optical phonon energies, which could suppress thermal backfilling of the lower laser level [19]. However, the epitaxy of these materials is more challenging and/or less mature than that of GaAs. Currently, the origins of inter-laboratory reproducibility issues remain elusive. These variations are likely attributable to differences in growth calibration protocols and criteria, as well as the stability and specific configuration of the MBE system. Tackling this challenge will require enhanced collaboration between laboratories, for example through round-robin growth experiments, to standardize protocols and refine AR designs through mutual feedback, thereby driving further performance improvements.

**Acknowledgements**

## 2.6 Heterogeneous Quantum Cascade Lasers

Michael Jaidl, Photonics Institute, TU Wien, Gusshausstrasse 27-29, 1040 Vienna, Austria; michael.jaidl@tuwien.ac.at
Karl Unterrainer, Photonics Institute, TU Wien, Gusshausstrasse 27-29, 1040 Vienna, Austria; karl.unterrainer@tuwien.ac.at

**Status**

Traditional quantum cascade laser (QCL) active regions typically consist of a series of identical, periodically repeated layers, which results in a high gain at the designed wavelength. However, this design limits the frequency range of the emitted light due to a restricted gain bandwidth. To overcome this limitation, heterogeneous QCLs were developed, incorporating multiple active region sections (sub-stacks) with distinct emission wavelengths into a single device. Each sub-stack is specifically designed to emit at a different wavelength, and they are combined in a stacked configuration, allowing for a broader and more tunable emission range. This approach leverages the scalable nature of QCLs, and various designs have been proposed and optimized theoretically to maximize performance [1]. Initially introduced for the mid-infrared region [2,3], heterogeneous QCLs have since expanded into the terahertz (THz) range [4-8]. Ultra-broadband emission in the mid-infrared and terahertz spectral region is advantageous for several applications. Spectroscopic applications such as chemical sensing and environmental monitoring benefit from the broad bandwidth since multiple absorption lines may need to be detected simultaneously. In medical and industrial imaging, these QCLs can enhance contrast by providing multi-wavelength illumination, improving the identification of different materials or biological tissues. The tunability and broad frequency range make heterogeneous QCLs attractive for future high-frequency telecommunication systems, especially in the terahertz range. Furthermore, their ability to operate across different wavelengths makes them useful in security for detecting hazardous substances and in defence for secure communication systems that rely on terahertz frequencies. Heterogeneous QC active regions can provide gain which spans over an octave. If it is possible to compensate for the dispersion over the emission bandwidth in a QCL, octave-spanning frequency comb formation is feasible. This would path the way for the full stabilization of the comb by utilizing the f-2f-technique and achieving frequency combs with high precision. Two promising candidates for this achievement are the heterogeneous active regions published by Rösch et al. [7] and Jaidl et al. [8]. The first one consists of three designs based on a hybrid bound-to-continuum/resonant phonon design and emits in the range of 1.64 THz to 3.35 THz, covering one octave. The maximum operating temperatures are 55 K in continuous-wave operation and 105 K in pulsed operation. The second one includes five active region designs based on a three-well resonant phonon design which was optimized for high temperature operation. Its emission spans 1.37 octaves, covering a bandwidth from 1.9 to 4.5 THz (Fig. 1 &2). Due to the operation temperature optimization, the maximum operating temperature could be increased to 58.5 K in continuous-wave operation and 143 K in pulsed operation.

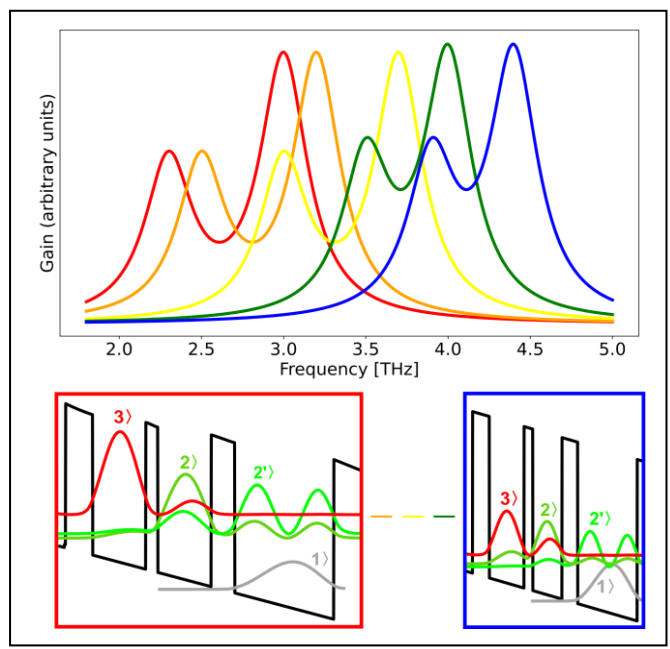

**Figure 1.** Gain cross-section of the five sub-stacks used in [Jaidl]. The double-peak results from two transitions occurring in each sub-stack. The lower panels show the band diagrams of the active regions with the highest and the lowest emission wavelength, respectively. The optical transitions occur between the upper laser level 3⟩ and the two lower laser levels 2⟩ and 2'⟩. The latter also functions as extractor level. Electrons are depopulated from the lower laser level to the injector level 1⟩ via the emission of a longitudinal-optical phonon.

**Current and Future Challenges**

Fabrication complexity remains a hurdle. The intricate multi-layer structure of QCLs requires advanced epitaxial growth techniques demanding precise control over each layer to ensure high-quality interfaces with minimal defects. This implicates the requirement of stable growth conditions over a comparatively long period of time since the growth of a heterogeneous active region takes a couple of hours.

A key point in the design of heterogeneous QC active regions is the matching of the currents flowing through the individual sub-stacks. Mismatches in current distribution across the different active regions can lead to discontinuities of the electric field. This results in misaligned energy levels and inefficient electron transport, increased heat generation, and uneven optical gain, which can impair performance and lead to undesired optical spectra. The operation temperature is an important issue especially for THz QCLs. So far, room-temperature operation has not been achieved in these devices. In order to pave the way for portable and commercial use of ultra-broadband heterogeneous QCLs, the requirement of cooling needs to be reduced.

Octave-spanning frequency comb formation in a QCL is a milestone to be reached yet. The greatest obstacle preventing this achievement is the nonlinear dispersion which hampers the emission of equidistant modes over the whole gain bandwidth.

**Advances in Science and Technology to Meet Challenges**

To tackle the above-mentioned challenges, several advances in different fields are necessary. Stable growth conditions and sophisticated feedback systems, such as in situ monitoring tools, are essential to ensure defect-free interfaces over extended growth periods. Thereby it is necessary to consider different growth conditions for changing layer thicknesses in heterogeneous active regions.

It is necessary to carefully engineer the current matching between sections, which can ensure that each active region contributes efficiently to the laser output, resulting in improved power efficiency, stability, and spectral control. Computational modeling and machine learning play an increasingly important role, helping to predict electronic and optical properties and optimize active region designs for target wavelengths and efficiency. A better understanding of the actual level alignment and consequential electron transport through the structure is crucial for the design of high-performance active regions [9].

To achieve room temperature operation in THz QCL, further research on active region designs is necessary. A promising candidate is the two-well design with a current record of T_max = 261 K [10]. This particular design would bring another advantage for heterogeneous active regions. Since one period of the structure is thinner compared to other designs consisting of three or more wells, a higher number of periods can be included in the heterogeneous structure. In this way it is possible, either to increase the gain for the individual sub-stacks or to include additional designs emitting at other wavelengths to broaden the gain bandwidth even further.

In order to realize an octave-spanning frequency comb, the dispersion needs to be tackled. A large contribution of the dispersion stems from the material, but also waveguide dispersion and gain-induced dispersion play a role. Due to the vicinity of the Reststrahlen band in THz QCLs, the material has a large impact on the dispersion towards higher frequencies. Suitable dispersion compensation techniques such as chirped/coupled waveguides [11,12], Gires-Tournois-Interferometer mirrors [13] or the implementation of graphene layers [14] are necessary to flatten the dispersion curve over the whole gain bandwidth. In this way, octave-spanning comb emission could be feasible.

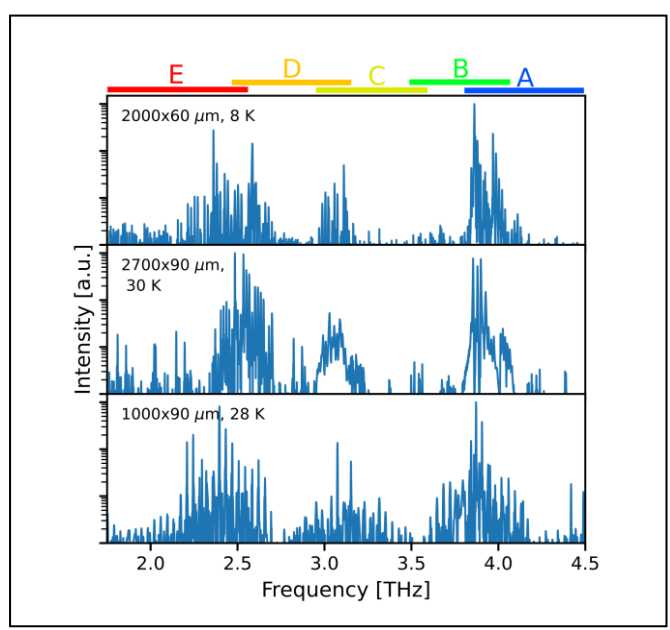

Figure 2. Optical spectra of three different devices in continuous-wave operation using a heterogeneous active region consisting of five sub-stacks. The gain bandwidth spans 2.6 THz from 1.9 to 4.5 THz, spanning 1.37 octaves. The coloured bars on top indicate the gain region of the individual sub-stacks. Reprinted from [Jaidl].

**Concluding Remarks**

Heterogeneous QCLs represent a transformative innovation in photonic technology, offering a broader and more tuneable emission range compared to traditional designs. These advancements open up new horizons in diverse fields, from environmental monitoring and medical imaging to telecommunications and security, where ultra-broadband and tuneable light sources are critical. The potential for octave-spanning frequency combs further underscores the promise of heterogeneous QCLs, enabling high-precision applications and advancing our understanding of quantum systems. The path forward for heterogeneous QCLs is both challenging and exciting, promising substantial technological and scientific breakthroughs. By addressing the existing limitations through concerted research and innovation, the realization of their full capabilities will mark a significant leap in photonics, benefiting a wide array of critical industries and applications.

**Acknowledgements**

The authors acknowledge financial support by the Austrian Science Fund FWF (TeraLearn P35932-N).

### 3.1 Broadband and frequency comb emission in quantum cascade lasers

Giacomo Scalari, Jérôme Faist, ETH Zürich
gscalari@ethz.ch, jfaist@ethz.ch

**Status**

The broadband operation and tunability of lasers has been a central theme since the early days of the field in 1960. Quantum Cascade Lasers (QCLs) have had in the frequency agility one of their main selling points. The same band curvature of the subband results in a symmetric (Lorentzian) density of states, featuring transparency on both sides of the central frequency. This feature has been largely exploited in order to engineer broadband gain regions, often based on multi-stacks sharing the same waveguide, as visible in Fig.1(a,b) [1]. In the Mid-IR, multi-stack active regions have been used for broadband single mode emission in external cavities [2] (see Fig.1(c-e)), and broadband QCLs have been combined in multi-chip external cavity devices for ultra-broadband operation present on the market since more than a decade and routinely employed for absorption spectroscopy and for more fundamental studies like near-field nanoscopy. In terms of pure optical bandwidth, the THz region greatly benefited from the multi-stack concept as the double metal waveguide features an almost frequency-independent figure of merit and unchanged overlap factor over a multi-octave frequency span. Following the multi-stack approach, octave-spanning emission from a single laser has been reported in 2015 [3]; this constitutes a first not only for QCLs but for semiconductor lasers in general. Broadband active regions have been successfully used for difference frequency generation of THz and sub-THz radiation from monolithic devices [4], [5] and even integrated with external cavities [6] covering an impressive 1.7-5 THz frequency interval. In terms of bandwidth coverage from monolithic devices, interesting results have been recently obtained by employing broadband active regions strongly driven with RF pulses highly detuned from the round trip (on the low frequency side): such devices reach 250 $cm^{-1}$ of bandwidth [7] and have been successfully used in combination with a fast Fourier transform spectrometer [8].

The strive for broadband and flat-top gain profiles in quantum cascade lasers is nowadays driven by the demand of broadband frequency combs and lasers for multi-frequency external cavity devices. After the demonstration of actively mode-locked Mid-IR and THz QCLs [9], frequency comb research in QCLs was ignited in 2012 by the discovery that these devices naturally mode lock into an FM state that minimizes amplitude variations ensuring at the same time stable phase-locked modes [10]. The following 10 years have seen an impressive development of this field, with substantial advancements in device performance and theoretical understanding [11], [12]. QCL-based frequency combs are nowadays spanning both Mid-IR and THz [13] frequency ranges: broadband FM combs over 100 $cm^{-1}$ at 1300 $cm^{-1}$ (7.7 μm) have been demonstrated[14] as well as bandwidths in excess of 1 THz at 3 THz central frequency [15].

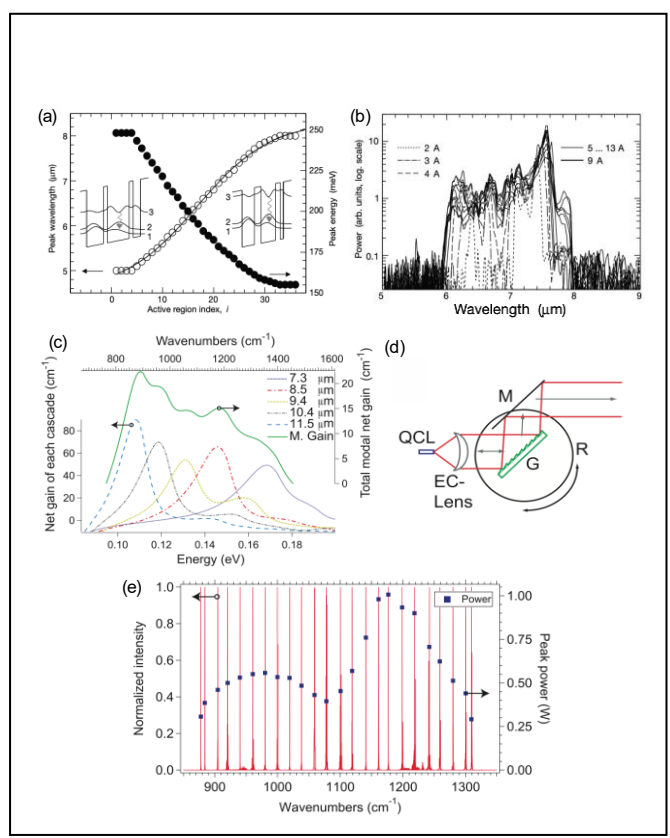

**Figure 1.** (a): Calculated peak wavelength (open circles) and peak energy (filled circles) versus active region index i. The solid curve is a least-squares fit using a Fermi-type function. The left and right inset shows a band structure diagram of a `three-well vertical transition'. (b): Above-threshold spectra of the supercontinuum laser. Spectra obtained at the various peak current levels indicated from a laser operated in pulsed mode at cryogenic temperature. (Adapted from Ref. [1]). (c): Simulated net gain including resonant absorption lossesvof individual cascades. All cascades are simulated at approximately the same current density of 6.5 kA cm$^{-2}$. The total modal net gain isvthe sum of all gains of the cascades times the modal overlap. (d): External cavity-QCL operated in the Littrow configuration. (e): Tuning behaviour with corresponding peak optical output power of the External cavity -QCL at various wavelengths. (Adapted from Ref. [2] )

**Current and Future Challenges**

The main challenge is surely represented by the emission bandwidth and the full exploitation of the natural broad electrical bandwidth of the devices: for many applications a broad optical bandwidth is essential in order, for example, to unambiguously identify chemical species via absorption spectroscopy. The electrical bandwidth is tightly connected to the possibility to control the quantum cascade broadband devices and combs with suitable radio-frequency signals. On the other side, integration of devices is of great value when targeting widespread applications where the laser is part of a more complex system including electronics and signal processing. The development of Photonic Integrated circuits in the Mid-Ir would offer great possibilities both for bandwidth broadening (i.e. supercontinuum generation via non-linear waveguides) as well as for on-chip integrated sensors and spectrometers.

**Advances in Science and Technology to Meet Challenges**

Recent advances in QCL frequency combs address some of the challenges mentioned above. Lasers based on ring resonators are seeing a rapid development: after the demonstration of self-starting solitons in circular structures[16], remarkable progress in terms of understanding and control of the solitons and their routing via integrated active waveguides has been demonstrated [17] (see Fig.2(a,b)), basically unifying the fields of on-chip Kerr microcombs and QCLs combs. Another advancement that shows very interesting potentialities is the recently demonstrated quantum walk comb [14]. Such device, operating in a low-defect ring cavity in absence of spatial hole burning, can be completely controlled via RF modulation, giving rise to an FM comb with spectral envelope that can be predicted analytically. Particularly, the emission spectrum can be further controlled and engineered by RF synthesis exploiting the peculiar properties of the ring resonators modes that can be coupled, via a suitable modulation M, in analogy to lattice sites of a synthetic space (see Fig.2(c-e)). In general, the availability of these and other on-chip ring comb sources [18] opens great opportunities for applications, for example in integrated dual-comb spectrometers. In the last 10 years, there has been an increasing effort in many fields of photonics towards the integration of III-V gain elements onto silicon photonics platforms. Also in the QCL community there has been quite remarkable progress in this direction: both hybrid integration and homogeneous integration have seen significant advances. The possibility to directly grow QCLs with good performance onto a silicon wafer [19] offers an extremely interesting option for Mid-IR silicon photonics. Homogeneous integration onto the InP platform with waveguides and evanescent couplers [20] and the concomitant demonstration of high quality factor ($Q>10^5$) Germanium ring resonators [21] holds promises for interesting developments in the field of mid-IR Kerr combs and more generally for a Mid-IR integrated photonics platform. Hybrid integration of QCLs on Silicon [22] has been as well demonstrated for both Mid-IR and THz QCLs, showing that several approaches can be used towards Mid-IR and eventually THz photonics circuits.

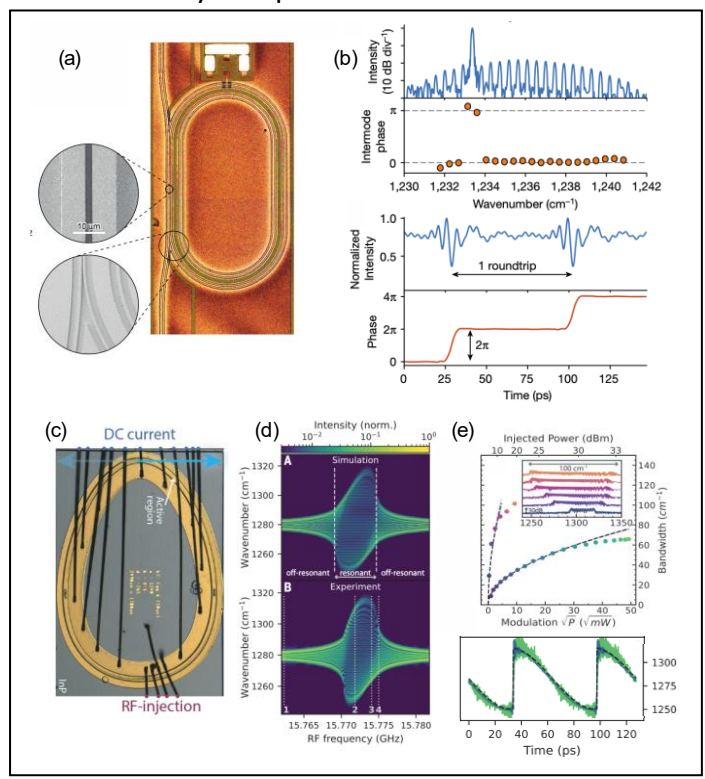

Figure 2. (a): Microscope image of the QCL ring and waveguide coupler, with separate electrical contacts. (b) spectrum with intermode phases, time reconstruction and phase of the emitted radiation. (Adapted from Ref. [17]). (c): microscope image of QCL egg-shaped resonator.(b): Spectra for sweeps of the injection frequency, as predicted by simulations based on the Maxwell-Bloch equations (A) and measured in FTIR (B). the progressive broadening of the spectra with increasing injection power (linear scale)] follows the predicted $\sqrt{M}$ dependence. Discontinuous phases at half the roundtrip frequency in the quantum walk resonant regime (adapted from Ref. [14])

**Concluding Remarks**

The capability to engineer the shape, position and width of the gain curve is a hallmark of the quantum cascade laser architecture. For this reason, it is not a surprise that the design of broadband devices and their operation as optical frequency comb remains a most active area of research. On one side, the very fast gain recovery time of the quantum cascade active region material allows a dynamical control of the comb state and opens new avenues in comb phase control and spectra shaping, defining a new research frontier for ultrafast physics. On the other the ability to engineer optical non-linearities and combine them monolithically with the source allow the development of fully integrated ultrafast devices for applications.

**Acknowledgements**

*The Authors acknowledge financial support from SNF project 200021-232335.*

## 3.2 Soliton quantum cascade lasers

L. L. Columbo[1], M. Brambilla[2,3], M. Piccardo[4,5,6]

[1]*Dipartimento di Elettronica e Telecomunicazioni, Politecnico di Torino, 10129 Torino, Italy*

[2]*Dipartimento di Fisica Interateneo, Università e Politecnico di Bari, 70126 Bari, Italy*

[3]*CNR-Istituto di Fotonica e Nanotecnologie, UOS Bari, 70126 Bari, Italy*

[4]*Harvard John A. Paulson School of Engineering and Applied Sciences, Harvard University, Cambridge, MA 02138, USA*

[5]*Department of Physics, Instituto Superior Técnico Universidade de Lisboa, 1049-001 Lisbon, Portugal*

[6]*Instituto de Engenharia de Sistemas e Computadores – Microsistemas e Nanotecnologias (INESC MN), 1000-029 Lisbon, Portugal*

[lorenzo.columbo@polito.it; massimo.brambilla@poliba.it; marco.piccardo@tecnico.ulisboa.pt]

**Status**

A peculiar dynamical property of Quantum Cascade Lasers (QCLs) [1] consists in a carrier's lifetime dominated by fast phonon scattering that makes QCLs class A lasers where the electric field is the slowest evolving variable [2]. In a standard Fabry-Perot (FP) configuration this favors the phenomenon of Spatial Hole Burning (SHB) in the carrier's density spatial profile. As a consequence, Optical Frequency Combs (OFCs) in standard FP QCLs are associated with a small Amplitude Modulation (AM) and linear frequency chirp as quasi Frequency Modulated (FM) combs. OFCs are of particular interest in the mid-IR and THz regions of the electromagnetic spectrum (from 1 THz to 100 THz) where radiation from QCLs is typically emitted for e.g. high-resolution molecular spectroscopy [3,4].

Recently two research groups at Harvard University [5] and at ETH Zurich [6] reported almost simultaneously experimental evidence of OFCs in an unconventional unidirectional ring configuration that naturally leads to integration in photonic circuits [7]. This new type of OFCs were associated with AM solitary (e.g. non-dispersive) structures with $sech^2$ comb spectral envelope. Physically, the mechanism of CW destabilization that leads first to multiwavelength emission and then to phase locking among the laser lines in presence of sufficiently efficient Four Wave Mixing is not ascribed to SHB, but to a non-zero Linewidth Enhancement Factor (LEF) that maps intensity modulations into phase (or frequency) modulations. This allows under realistic working hypotheses to describe the complex multimode dynamics of unidirectional ring QCL dynamics in terms of a single order parameter equation for the electric field that has the form of a Complex Ginzburg Landau Equation (CGLE) [8]. This evidence puts QCLs in the same class of universality of a wealth of nonlinear extended systems in quantum-physics, hydrodynamics, chemistry, etc. Thus, in the general framework of the CGLE the solitary structures associated to OFCs in unidirectional rings QCLs were interpreted as *homoclons*, i.e. shallow bright solitary pulses emerging from the phase turbulence regime of the CGLE [5] or *Nozaki-Bekki solitons*, i.e. dark solitary holes coexisting with stable CW solutions [9,10].

The analogy with the bright Kerr solitons/combs reported in passive, coherently driven on-chip high-Q microring configurations [11], led to the conception of a novel device represented by a unidirectional racetrack QCL driven by another laser, where the existence of bright (high contrast) solitons were first predicted and then experimentally observed (see Fig. 1) [12,13].

More complex configurations where two or more ring QCLs are coupled together to produce new hybrid self-organized states can be also envisaged [14].

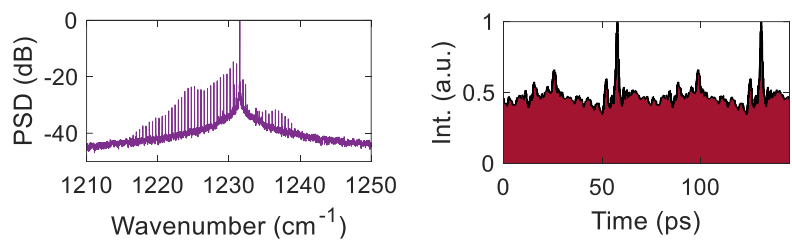

Fig. 1. Experimental optical spectrum (left) and intensity trace (right) of two stable bright solitons (one per cavity roundtrip) traveling in an optically driven QCL (parameters as in Fig. 4 of [13]).

**Current and Future Challenges**

Solitonic emission in above-threshold QCLs offers the twofold opportunity of achieving short, high contrast and possibly bright pulses and simultaneously exploiting robust OFC associated to their spectrum. This opens the door for applications to on-chip soliton generation in the mid-IR for e.g. supercontinuum generation, broad-band free space optical communication, high resolution spectroscopy.

A primary interest resides in achieving short duration and high soliton power. Sub-ps pulses are achieved in driven, passive Kerr microresonators [15] while in mid-IR QCLs, *Nozaki-Bekki solitons* [9] and bright ps pulses have recently been obtained [10,13]. Further timescale reduction might encompass both better control of the medium fast dynamics - the QCL interband coherence relaxation time being fundamental for soliton scales {12} - and possibly dispersion engineering in the QCLs [16, 17]. Currently, power increase techniques include generation-stage control: pump optimization, gain stage design, device fabrication/mounting and thermal management (see next section). Also, downstream amplification of the bright soliton will be a key to enhance both its intensity and its contrast, by filtering out the CW background [13].

Being self-confined wave packets, solitons are an obvious means to encode/process information in a coherent beam. Instrumental to such goal is the possibility to address solitons deterministically turning them on and off, the control of their location in the resonator and the control of mutual interactions, leading to formation of soliton molecules and trains, whose spectral properties immediately appear attractive in terms of OFC properties [18].

Such features were achieved acting on global laser parameters (e.g. QCL detuning or bias current, in lasers [13] or in passive Kerr microresonators [19]) as opposed to spatially distributed features in the device.

The control of solitons through external "channels" (driving field intensity/phase or bias profile/dynamics) is of paramount relevance to soliton encoding and OFC engineering: clear evidence was initially offered in fiber lasers [20] but meeting the challenge to achieve short modulation times and high rates, necessary to address mid-IR QCLs, is an indispensable step ahead.

A different control pathway for fast soliton addressing might reside in the phase relations the solitons exhibit e.g. in the Argand plane, where it most often shows a looping orbit around a fixed point (possibly, the origin, thus revealing a topological charge) [21]. In fact, theoretical/numerical evidence shows that solitons can be excited by superimposing a (0-2π) phase kink to the homogeneous input field.

Of course, the OFC associated with the deterministically controlled soliton train/molecule will change its shape and peak separation (e.g. as in trains of N equally spaced solitons, N being controlled by the external addressing) [12].

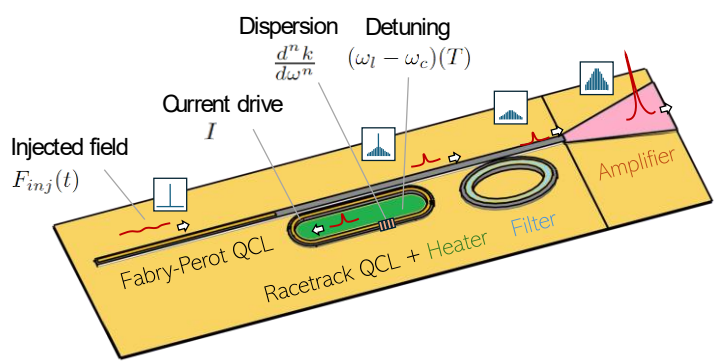

Fig. 2. Schematic of the envisioned integrated chip for high contrast, ultrashort, high energy QCL soliton generation for mid-IR photonics.

**Advances in Science and Technology to Meet Challenges**

On the frontiers of soliton generation, management, and control, significant theoretical progress in QCLs has been made, but major advances are still needed. Specifically, increasing soliton power, reducing pulse duration, and achieving precise soliton control as proposed in the previous section require substantial technological advancements (see Fig. 2 for an artistic view of a QCL soliton generation photonic chip).

Enhancing soliton peak power requires optimization of QCL design, focusing on improved thermal dissipation via epitaxial-side-down mounting and gold plating. Such optimizations could increase soliton power by an order of magnitude. Additionally, the integration of semiconductor master oscillator power amplifiers alongside the QCL promises efficient amplification, with demonstrated factors of up to 100 in ridge waveguide geometry, promising to boost soliton peak to boost soliton pulse energy in the 100 pJ range. Advances in high-power gain medium design, particularly through strained-balanced active regions, could lead to emission levels exceeding several Watts, reducing the amplification requirements for achieving high soliton peak energies [22].

To achieve shorter soliton pulse durations, technological developments aimed at increasing the spectral bandwidth are necessary. Spectral engineering of the QCL active region can facilitate broader gain spectra, which are key to reducing pulse duration below the current demonstrations of 1 ps. Furthermore, on-chip supercontinuum generation using tapered nonlinear waveguides, such as those made from InGaAs, could offer a promising route for spectral broadening. Amplified pulses can be propagated through passive nonlinear waveguides to produce output pulses in the sub-ps regime, enabling ultrafast applications in the mid-IR.

Soliton management, including writing and erasing solitons, remains an open area. To date, experimental seeding of solitons in ring QCLs with external pulses has not been demonstrated. An approach could involve an array of injector Fabry-Perot QCLs, actively mode-locked to produce pulses in the range of 1-10 ps. With active RF synchronization, these injector pulses could be precisely injected into the ring QCL to write and erase solitons as needed, providing a flexible mechanism for soliton manipulation.

In a fully integrated injector-ring configuration, this would create a reconfigurable, stable on-chip soliton system, significantly broadening the potential applications of mid-IR photonic technologies.

**Concluding Remarks**

The evidence of solitons in QCLs reported almost 5 years ago in an unconventional unidirectional ring configuration triggered an increasing interest in both their fundamental nature as self-localized dissipative structures and their spectral content as AM optical frequency combs opposed to the FM modulated characteristic of standard FP configuration.

Moreover, the description of soliton formation in QCLs through the universal CGLE allowed to foresee some of their peculiar properties exploiting the analogy with other physical nonlinear extended systems.

For example, the advancing focus on soliton management in QCLs is presently harvesting benefits from the extended knowledge amassed in over three decades of research on dissipative solitons in nonlinear resonators of the most varied flavors. Theoretical indications on soliton existence and robustness in practical operational regimes, soliton duration, power, addressing and composition, are a primary objective. The former guidelines will need matching the open and challenging technological and experimental feats. Using nonlinear dispersion engineering and optical amplification, preliminary experimental results let conceive the realization of the next generation of integrated sources of externally addressable sub-picosecond solitons with pulse energy in the 100 pJ range for applications to mid-IR photonics.

**Acknowledgements**

We thank the research groups of Harvard (D. Kazakov, T. Letsou, F. Capasso), TU Wien (B. Schwarz, N. Opacak. S. Dal Cin) and Università dell'Insubria in Como (F. Prati, L. A. Lugiato) for collaborations leading to several ideas referenced in this perspective.

## 3.3 Pulse generation from Quantum Cascade Lasers

S. S. Dhillon
[1] Laboratoire de Physique de l'Ecole Normale Supérieure, ENS, Université PSL, CNRS, Sorbonne Université, Université de Paris-Cité, 24 rue Lhomond, 75005 Paris, France
sukhdeep.dhillon@ens.fr

**Status**

The generation of ultrafast and intense light pulses is an underpinning technology throughout the electromagnetic spectrum enabling industrial and academic applications across the sciences, from frequency comb (FC) spectroscopy to ultrafast biophotonics. An important system has been semiconductor-based devices for pulse generation in the optical range. These benefit from inexpensive wafer scale production and low system costs. However, in the mid-infrared (MIR) and THz range, spectral ranges with countless molecular signatures and transparent atmospheric windows, pulse generation from QCLs has lagged compared to their semiconductor counterparts in the near-infrared range. This is mainly down to the unique dynamics of QCLs where the gain relaxation is considerably faster than the photon round trip time. This unfortunately prevents the build-up of high peak power pulses in the cavity when used in conjunction with, for example, saturable absorbers, but nonetheless offers the possibility of ultrafast modulation of the gain. This has opened up interesting approaches for pulse generation from these electrically pumped sources, as well as understanding ultrafast laser physics in semiconductor devices. These developments have the potential to bring new approaches in molecular sensing as well as investigating dynamics of the many ultrafast processes that occur in this spectral region, or new applications in microwave photonics for the generation of low-noise microwave signals.

Although frequency comb (FC) operation has been extensively demonstrated in QCLs, these FCs are based on a third order nonlinearity and are therefore inherently frequency modulated [1,2] i.e. these do not generate pulses. Nonetheless, there has been considerable advances in pulse generation in QCLs. Although pulse generation was already being investigated in the late 1990's, it took until the late 2000's to categorically show pulse generation, first in MIR QCLs[3] and then in the THz[4]. A range of approaches have been applied, including specially designed bandstructures for slower dynamics (but lower performance), and a strong RF modulation to fix the phases and mode spacing[5,6]. Recent work has shown femtosecond generation through pulse compression using an external geometry[7], distributed two dimensional saturable absorbers for passive modelocking in the THz[8], soliton pulse operation in ring cavities [9] and gain switched QCLs[10]. Of relevance to the MIR, significant advances have also been made in interband cascade lasers (ICLs), with similar approaches to that of QCLs, in particular to a strong RF modulation to force the laser to generate pulses[11].

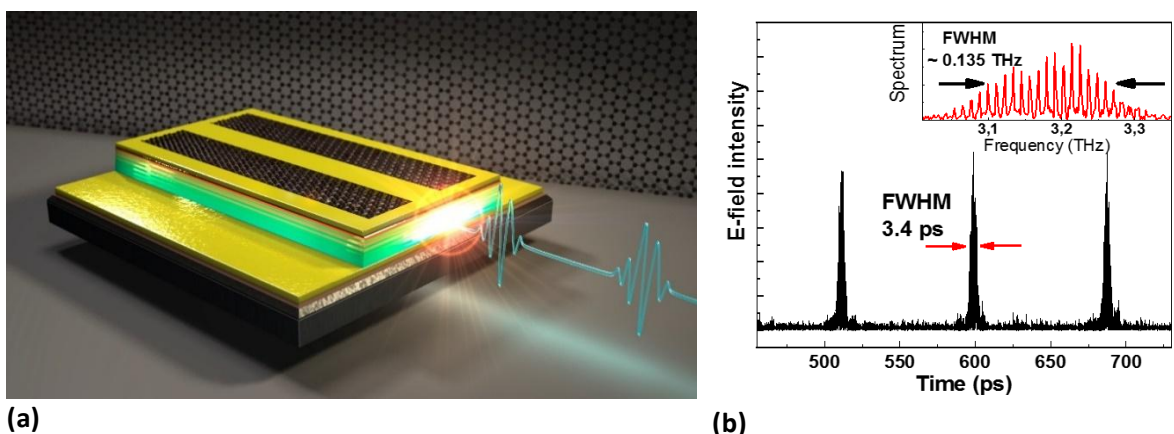

**Figure 1.** a) Schematic of pulse generation from a THz QCL with distributed saturable absorber. b) 3.4 ps pulses generate from a THz QCL operating at 3.2 THz using active modelocking and dispersion compensation

All these approaches have gone hand-in-hand with developments of time resolved theoretical tools[12,13] and new approaches for ultrafast characterisation of QCLs that was lacking previously. For the latter, this has led to a range of developments, from techniques based on coherent detection using sampling techniques[5], dual comb[14], two photon detection[3], up-conversion methods[7], to SWIFT spectroscopy[2] etc that have all permitted to categorically demonstrate pulse generation in QCLs and ICLs.

**Current and Future Challenges**

Despite these new approaches for research in QCL pulse generation and their ability to be engineered to operate in novel modelocking regimes, the translation of these ultrafast devices to widescale use outside a research environment has still to be realised. Immediate challenges include the development of easy-to-use systems, that could benefit to both academia and industry, as has been done other spectral regions. Underpinning these objectives is the need to reduce the complexity of certain approaches as well as reducing the high RF power requirements in the case of active modulation for pulse generation. Beyond sources, the intricacy of ultrafast detection needs to be addressed such that it is more democratic in terms of measuring and determination of pulse generation in the MIR and THz. Here, the parallel developments of sensitive, high-speed detectors would also considerably advance the field, especially in the THz range.

Beyond engineering of pulsed QCL systems, another challenge is that pulses are typically of low peak power and duration in the picosecond region. There is therefore a requirement for complementary devices to increase power and spectral bandwidth. As these systems are developed for shorter pulses, methods to control the index dispersion will become vital. Approaches to develop on-chip high peak power with large power per mode, operating across the entire dynamic range of the laser, remains a topic of considerable current interest, as well as being able to realise low repetition rates for certain spectroscopic applications or the targeting of high peak powers beyond on-chip solutions. (Currently the repetition rates are typically large owing to the Fabry-Perot cavities used). Pure passive modelocking (i.e. without an active modulation) remains a challenge despite advances in the THz region. The realization of saturable absorbers for the MIR and THz region is also challenging but would benefit the realisation of shorter and more intense pulse generation in the case of devices with relatively long carrier dynamics.

**Advances in Science and Technology to Meet Challenges**

The advances required will depend on the type of geometry and if it is on-chip or external cavities. Regarding on-chip geometries that are typically limited in output powers, and where higher peak powers could be achieved through a MOPA type geometry[15], coupled either to a RF modulated Fabry Perot or soliton-based ring type cavity, where the pulse generation can be performed separately to an amplifier section. External cavities or geometries that have been developed recently for the MIR, could be adapted for the THz that would require the development of low loss components such as adapted coatings and gratings that remains challenging. This will lose somewhat the benefits of a compact system compared to on-chip systems but would benefit in terms of fine control of dispersion.

In terms of bandwidths, dispersion compensation can be controlled through on-chip geometries including chirped[2] or GTI [16]structures although the latter can be limited in bandwidth. For pulse generation with repetitions rates in the GHz region, further considerations of waveguides that simultaneously guide RF and the MIR/THz frequencies would be beneficial where the interaction of RF and the optical mode has shown to effect the generation of electrical beatnotes [17]. In terms of applying saturable absorbers for the MIR distributed along the laser cavity, as has been shown in THz QCLs [8], recent developments in 2D materials could be equally applied to MIR QCLs[18]. Considerations of extremely fast saturable lifetimes will be even more important here. As well as graphene, a range of TMDs have been discovered that show low saturation intensities and fast dynamics in the MIR and can be integrated on top of QCL ridges. A fully monolithic approach could be combining a QCL with an electrically controllable interband transition in the QCL claddings (in for example in Sb based system quantum wells[19]). Finally, new saturable absorbers/mirrors based on 2D materials[18] or controllable ultrafast light-matter coupling concepts[20] could be applied to ICLs where the dynamics, although still fast, are expected to be similar to slower than MIR QCLs. Finally, optically pumped external cavities of structures with MIR interband or interlevel transitions (hence slower overall dynamics) could be revisited taking into account advances made in the optical range in terms of peak power generation[21]. Finally recent ultrafast detectors in the THz range based on field effect transistors [22] could be used to facilitate pulse detection for THz QCLs, and could be applied to time-of-flight measurements.

**Concluding Remarks**

Pulse generation in QCLs has shown large developments over the last decade, with new insights that have permitted to understand and apply the role of the ultrafast dynamics for pulse generation despite the extremely short lifetimes in these electrically pumped devices. The current advances offer interesting perspectives and directions to take, although there a real challenges to bring these demonstrations outside of the lab and to fully understand pulse formation in these inherently ultrafast lasers. For the moment pulse generation is limited in terms of peak powers compared to femtosecond solid state or fiber lasers that have been used to generate MIR and THz pulses through difference frequency generation. However, there are real possibilities to go further with QCL based structures, possible by increasing pulse energies by more than an order or two of magnitude, as well as generating shorter pulses on the few cycles regime. There could be further gains by going to an external cavity arrangement for a compromise between repetition rate and peak powers. This could open interesting avenues for enhanced nonlinearities for supercontinuum generation or even broadband THz generation via difference frequency generation. These sources could also then by used to looked at dynamics of fundamental interactions in the MIR and THz region, as routinely performed in the visible or near-infrared, using the QCL philosophy of bandstructure design to precisely target material transitions from molecular to phonon dynamics and nonlinearities. Further the impact of these new ultrafast methods in the MIR and THz ranges will permit this spectral region to be further exploited. This would also open up new opportunities in the application of such sources beyond spectroscopy, non-destructive testing or dynamic studies. For example, difference frequency generation for low noise THz/RF generation would become of interest and where the quantum efficiency of converting MIR photons to the THz or GHz region would be an order of magnitude than that in the NIR.

## 3.4 Modelling of quantum cascade laser frequency combs

Carlo Silvestri

c.silvestri@uq.edu.au, School of Electrical Engineering and Computer Science, The University of Queensland, Brisbane, QLD 4072, Australia

Current address: Institute of Photonics and Optical Science (IPOS), School of Physics, The University of Sydney, NSW 2006, Australia

**Introduction**

First demonstrated in 2012 [1], self-starting quantum cascade laser (QCL) optical frequency combs (OFCs) are of great importance for applications such as high-precision spectroscopy and free-space optical communication [2]. Beyond their technological relevance, they also offer significant scientific interest due to their remarkable properties. These include the coexistence of amplitude and frequency modulation, with a linear chirp of the instantaneous frequency reported in mid-IR Fabry–Perot (FP) QCLs [3], and their emergence from a regime of phase turbulence in the ring configuration [4]. Another distinctive feature, observed experimentally in both mid-IR and terahertz (THz) QCLs, is the self-starting emission of harmonic frequency combs (HFCs), where the comb spacing is an integer multiple of the laser cavity's free spectral range [5]. These behaviours have been investigated from a theoretical perspective and reproduced through numerical simulations using a variety of models. In this chapter, we provide an overview of the main modelling approaches used to simulate and understand the physical properties of QCL combs, and we outline the key challenges in this field.

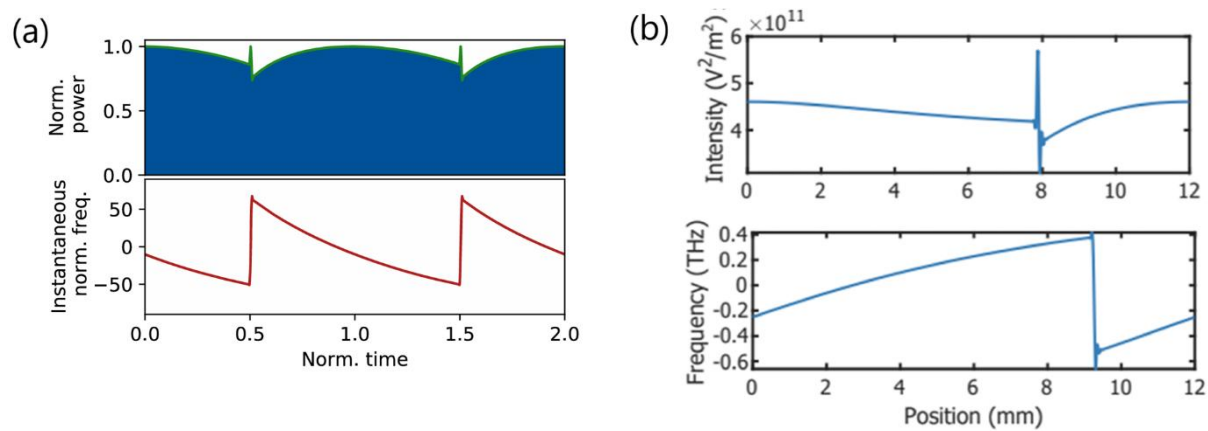

**Figure 1.** **(a)** Temporal evolution of the optical power (top) and instantaneous frequency (bottom) of a mid-IR Fabry–Perot QCL frequency comb, reproduced using the model presented in Ref. [14]. **(b)** Intracavity intensity (top) and instantaneous frequency (bottom) for the same class of devices, as reported in Ref. [15]. Figure (a) is reproduced from [14], and Figure (b) is reproduced from [15].

**Status**

A first class of models includes the so-called full models, such as the Maxwell–Bloch equations (MBEs) [6] and the effective semiconductor Maxwell–Bloch equations (ESMBEs) [7, 8]. MBEs have been employed to investigate the destabilization of single-mode emission that leads to multimode operation and comb formation in QCLs [9, 10], as well as to theoretically predict and interpret time-domain properties—such as the frequency-modulated (FM) behaviour that characterizes QCL combs [11]. Initial results revealed a pseudorandom trend in the instantaneous frequency which, although indicative of frequency modulation, did not correspond to the linear chirp later observed in experiments [11].

Concerning the multimode instability, theoretical studies on Fabry–Perot (FP) cavities based on MBEs—including carrier grating effects—have shown that spatial hole burning (SHB) leads to near-threshold multimode operation, in agreement with experimental observations [9, 10]. However, since there is no spatial hole burning (SHB) in the unidirectional ring configuration, frequency combs are observed only at pump levels approximately nine times above threshold in conventional ring MBEs, because of the Risken–Nummedal–Graham–Haken instability. Nevertheless, experiments have revealed comb formation close to threshold even in the ring configuration [4]. This suggests the presence of an additional mechanism responsible for destabilizing single-mode emission: the linewidth enhancement factor, or $\alpha$ factor, a characteristic parameter of semiconductor lasers that couples the phase and amplitude of the electric field [2]. Consequently, a new full model incorporating this effect—known as the effective semiconductor Maxwell–Bloch equations (ESMBEs)—was adopted to study first the ring configuration [8] and later extended to the Fabry–Perot case [12]. This model was derived from a phenomenological expression for the optical susceptibility of the semiconductor material. The ESMBEs enabled the reproduction of typical time traces and spectra in both the mid-IR and THz ranges, capturing key behaviors observed in experiments—such as linear chirp and harmonic comb formation [12, 13].

Although full models are well suited to reproducing experimental results, their mathematical complexity often makes gaining physical insight into certain phenomena less straightforward. This has motivated the development of reduced models based on a smaller set of equations.

A reduced model introduced in [14], derived directly from the full Maxwell–Bloch equations using a transfer-function approach, provided a deeper understanding of the phase dynamics behind QCL comb formation and highlighted the crucial role of both GVD and Kerr nonlinearity in establishing the FM comb regime. An example of temporal profile of a QCL comb exhibiting linear chirp and quasi constant intensity, reproduced with this model, is shown in Fig. 1(a).

Subsequently, a mean-field model developed for Fabry–Pérot QCLs reproduced FM combs through analytical "extendon" solutions of a nonlinear Schrödinger-type equation with a phase-induced potential [15]. These solutions describe linearly chirped, quasi-constant-intensity combs (see Fig. 1(b)), although the framework is more general: under ring-like boundary conditions it also predicts soliton states.

More recently, a reduced model based on an order-parameter approach has allowed unifying the dynamics of ring and FP cavities within a single generalized complex Ginzburg–Landau equation (CGLE) [16]. In the FP case this is a modified CGLE for an auxiliary field featuring a nonlocal integral term, which accounts for the interaction between the counter-propagating fields. For ring cavity, this term vanishes and the model reduces to the conventional CGLE, in agreement with [4].

The ring configuration has been extensively investigated within the framework of the complex Ginzburg–Landau equation (CGLE) [4, 17]. In the free-running regime, the CGLE predicts the existence of bright amplitude-modulated structures in the form of homoclons, which appear as islands of stability within the phase-mediated turbulence regime. In addition, it enables the prediction of dark pulses in the form of Nozaki–Bekki solitons [17]. Both types of localized structures, associated with frequency-comb formation, have been experimentally demonstrated [17].

In the presence of an injected field, the CGLE acquires a driving term and becomes equivalent to a generalized Lugiato–Lefever equation, capable of reproducing the dynamics of both active and passive optical cavities. Within this model, bright solitons have been predicted, and their existence has been experimentally verified [18].

**Current and future challenges**

One of the still puzzling open theoretical questions in the study of QCL frequency combs concerns the self-starting formation of harmonic combs. These regimes have been experimentally observed by several groups in both the THz and mid-IR spectral regions, and in both Fabry–Pérot and ring configurations [2, 5]. They have also been numerically reproduced using both full and reduced models. Studies based on Maxwell–Bloch equations for a two-level system showed that, in a Fabry–Pérot laser, when the single-mode emission becomes unstable, harmonic combs can be associated with peaks in the nonlinear parametric gain shifted by integer multiples of the cavity free spectral range [2]. More recently, their origin has been linked to a resonance between a cavity mode and an intrinsic oscillation frequency of the medium variables (carrier density and polarization). This frequency scales with the square root of the intracavity intensity—an established fingerprint of the Rabi frequency in coherent light–matter interactions—and has therefore been termed the Effective Rabi Frequency [19].

Another active line of theoretical research focuses on modelling experimental configurations designed to manipulate comb emission using external tools such as active and passive mode-locking, optical injection, and optical feedback. In ring QCL cavities, it has been shown that optical injection enables the generation of cavity solitons via an approach analogous to that employed in passive near-infrared microresonators [20]. Although this analogy has been formalised, its full theoretical exploration remains largely uncharted.

With regard to optical feedback, recent theoretical results have shown that it is possible to discretely tune the comb spacing and induce harmonic combs on demand [21], potentially opening new opportunities—particularly for sensing applications. A first step in this direction has been the theoretical investigation of the impact of frequency-dependent target reflectivity on comb emission when light is reinjected into the laser cavity.

Furthermore, another configuration currently considered in modelling is that of QCLs with a saturable absorber, following the experimental demonstration of pulse generation in a mode-locked THz QCL exploiting graphene as the absorber [22]. Recently, a novel reduced model accounting for a distributed saturable absorber in a ring cavity has theoretically predicted the formation of dissipative bright solitons [23]. These differ from those observed in injection-locked ring configurations in that they exhibit contrast equal to 1. This work conceives a novel compact, self-sustained ring configuration to generate localized structures in the form of solitary waves and thus open the doors to several applications in the field of Mid-IR and THz photonics, such as free-space communications and high resolution molecular spectroscopy.

**Concluding Remarks**

The development of theoretical models describing the dynamics of frequency combs in QCLs has progressed in close parallel with experimental advances. At present, some of the most active research directions include the study of ring cavities, with a particular focus on the generation of localized structures, as well as the manipulation of comb states through optical feedback or external control techniques such as RF injection.

The investigation of localized structures, in particular, establishes a strong connection with the physics of microcombs in passive resonators and offers a pathway for translating near-infrared techniques into the spectral regimes relevant to QCLs, namely the THz and mid-infrared. Optical feedback, on the other hand, is of significant interest for sensing applications and multispectral imaging techniques.

**Acknowledgements**
The author thanks Xiaoqiong Qi, Thomas Taimre and Aleksandar D. Rakić for fruitful and interesting discussions.

## 3.5 Broadening quantum cascade laser combs with loss- and gain-shaping

Mithun Roy[1,2] and David Burghoff[1,3]

[1]*Chandra Department of Electrical and Computer Engineering, Cockrell School of Engineering, The University of Texas at Austin, Austin, Texas 78712, USA*
[2]*mithunroy177@utexas.edu*
[3]*burghoff@utexas.edu*

**Status**

Since the first demonstrations of their operation as *spontaneous* frequency-comb generators [1], [2], there has been a continuous push toward developing broader quantum cascade laser (QCL) combs, as broad bandwidths are essential for most comb applications. For example, many chemical compounds have distinctive "fingerprints" in the mid-infrared and terahertz (THz) regions [Fig. 1(a)], and broad bandwidth would enable the detailed analysis of their complex spectra, improving the identification and quantification of the substances. QCL combs were found to prefer an interesting frequency-modulated (FM) regime [3]—a phenomenon that eventually was discovered in many other laser platforms [4], [5], [6]. However, even though gain media can be extremely broadband, even octave-spanning, QCL combs have never conclusively reached the bandwidths available above the threshold. This is despite extensive innovation in dispersion engineering [1], [2], [7], [8], cavity design [9], and novel geometries like rings [10].

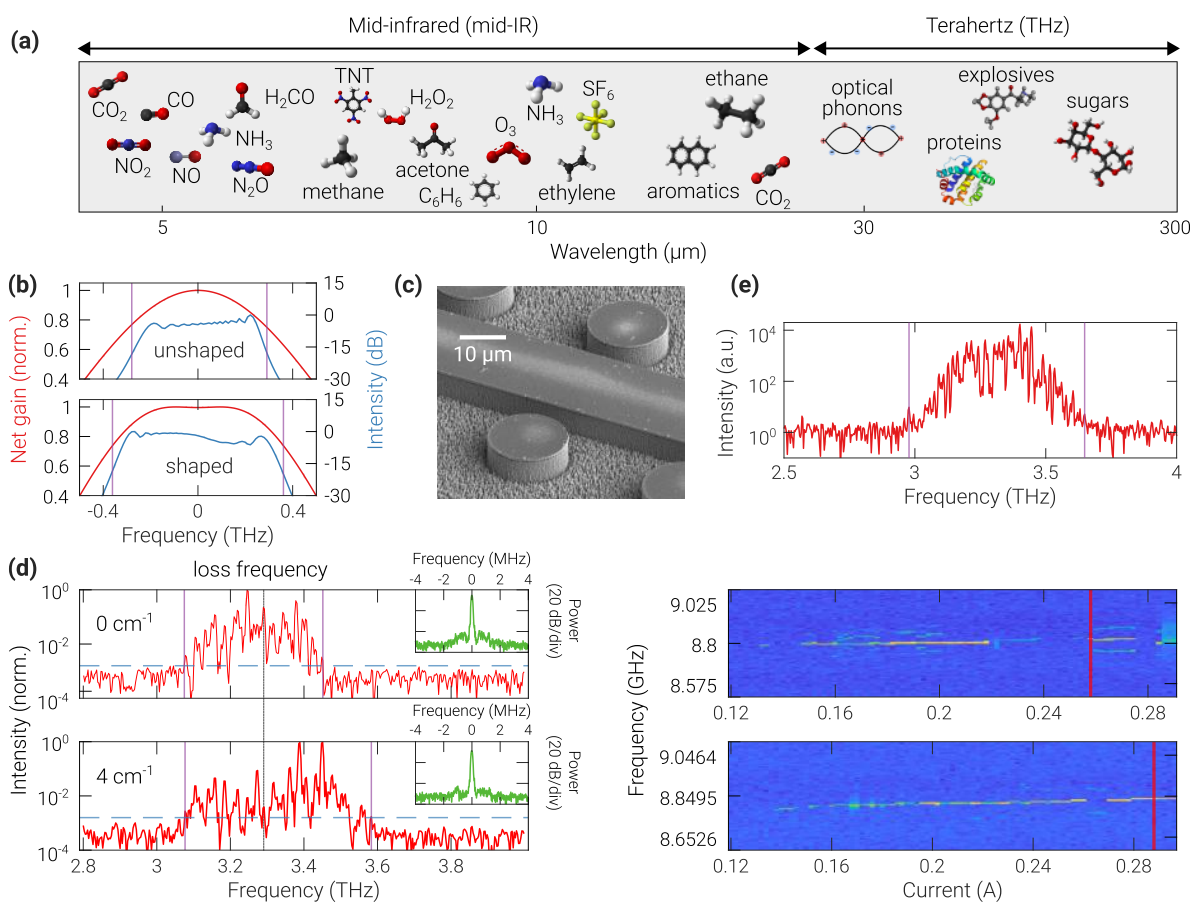

Figure 1. (a) Spectral fingerprints of chemical compounds [16]. (b) Mean-field simulation showing the effect of loss shaping on the bandwidth of a THz QCL comb. (c) SEM image of a loss-shaped device. (d) Broadest spectra, beatnotes, and beatnote maps for the reference FP and 4-cm$^{-1}$ device. (d) Spectrum of the broadest comb measured with a superconducting bolometer. Reproduced from [12].

By generalizing the Lugiato-Lefever equation, a mean-field theory was previously formulated to describe FM combs in Fabry-Perot (FP) lasers [11]. Without gain curvature, one can achieve an infinite bandwidth for the fundamental FM comb—referred to as an extendon, as it must extend throughout the cavity—by simply reducing the dispersion to zero. However, in the presence of gain curvature, even when the laser is far above the threshold and has ample gain across a broad bandwidth, there is no guarantee that the FM comb will be able to utilize all this gain. Given a value of gain curvature, *increasing* the sweep bandwidth of an extendon (e.g., by reducing dispersion) will *increase* the amount of amplitude modulation. Since an extendon is stable when its amplitude modulation is small, introducing a significant amount of amplitude fluctuation will destabilize it. This fluctuation can be reduced by lowering the gain curvature. Therefore, a flat gain will ensure that the extendon remains stable and can fully exploit the gain bandwidth. The major challenge with any FM comb is the gain flatness—small changes in the shape of the gain spectrum can lead to enormous changes in comb shape.

**Current and Future Challenges**

To counteract gain curvature, the concept of *loss shaping* was demonstrated [12], in which resonant loss was introduced with the peak-loss frequency aligned to the unsaturated gain center frequency. In Fig. 1(b), a generalized mean-field equation [12] was used to show the effect of such resonant-loss shaping on a THz QCL comb. Under optimal loss parameters, the theory predicts an enhancement of ~27% in comb bandwidth. To introduce resonant loss, small disks (or pillars) were fabricated on both sides of the cavity, covering almost the entire length of the cavity [Fig 1(c)]. Each of the disks is, in fact, the QCL gain medium in a waveguide. Light evanescently couples into these cavities, resulting in a resonant loss that can be controlled by varying the disk parameters. Loss shapers with 2, 4, and 8 $cm^{-1}$ peak loss, including identical and non-identical disks, were tested. The 4-$cm^{-1}$ devices, with identical and non-identical disks, demonstrated a significant increase in comb bandwidth (about 200 GHz or ~40%) as well as cleaner beatnote maps (narrow, single beatnotes essentially across the full dynamic range) compared to the reference FP [Fig 1(d)]. The coherence of the broadest-bandwidth comb was verified using SWIFTS [13]. The broadest comb's spectrum, measured with a superconducting bolometer, spanned ~700 GHz [Fig. 1(e)], ~80% of the lasing range of an identical wafer [14], showing loss shaping's effectiveness.

Large group velocity dispersions (GVDs) are detrimental to comb operation, but recent theories [11], [15] suggest that a minimum GVD value is necessary for comb stability. Figure 2(a) illustrates the interplay between GVD and gain curvature for a typical THz QCL comb, where the mean-field theory was employed to investigate the influence of these parameters on comb stability. The comb was considered stable if the coherence exceeded 0.999. The optimum dispersion is shown to vary quadratically with gain curvature. For GVD values higher than the optimum dispersion, stable and coherent combs are observed. In contrast, lower dispersion values lead to severe amplitude fluctuations, resulting in chaotic multimode behavior. This interdependence suggests the necessity of integrating a tunable dispersion compensator into the laser cavity for further optimization of comb bandwidth, as opposed to the constant compensator (double-chirped mirror) used so far. The losses introduced by the disks primarily arise from the radiation of coupled light; however, intersubband losses inherent to the gain medium also contribute. Next-generation loss shapers must consider intersubband loss frequency dependence near lasing to optimize performance. One could further flatten the gain by biasing the disks (which were not biased in [12]) to add resonant gain off the center frequency with minimal added complexity.

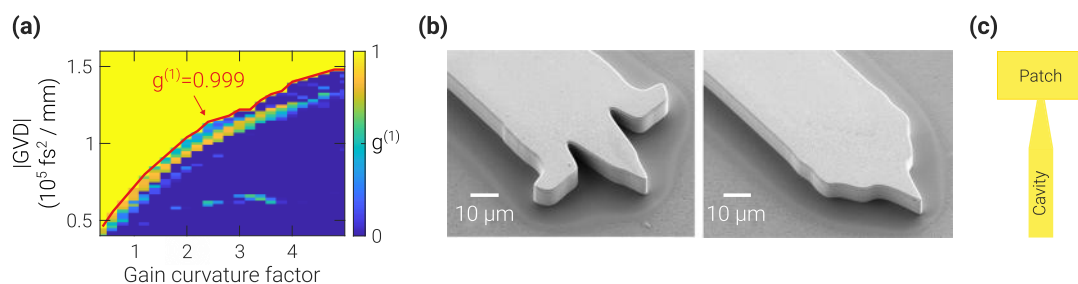

Figure 2. (a) Comb coherence ($g^{(1)}$ coherence) for a THz QCL. (b) SEM images of facet reflectors with different facet reflectivities. (c) Schematic diagram of a patch-antenna coupled QCL. Reproduced from [12], [17].

**Advances in Science and Technology to Meet Challenges**

The loss-shaping strategy utilizes disks to introduce resonant loss, although any resonant structure could be employed. A limitation of this design is that the distribution of disks along the cavity scatters the fundamental mode into higher-order transverse modes. While this issue can be mitigated through careful optimization of the inter-disk spacing and the Fabry-Perot (FP) cavity width, a better approach may involve integrating the loss-shaping mechanism at a cavity facet. For mid-infrared QCLs, loss shaping can be achieved through facet coatings that act as a notch filter. However, coatings are impractical for terahertz (THz) QCLs due to inefficient light outcoupling from subwavelength lasers. Instead, facet reflectors similar to those described in Ref. [16], could be designed [Fig. 2(b)]. Using an inverse-design approach, the precise facet shape can be optimized to enhance a specific figure of merit. This method not only addresses the loss-shaping requirements but also increases the outcoupled power compared to conventional cleaved facets. Alternatively, an antenna structure [Fig. 2(c)] could be designed to selectively radiate a portion of the spectrum centered around the gain-peak frequency. A significant challenge in disk biasing arises from the top metal layer, which, without adequate support, would become fragile and exhibit poor adhesion to the disks. The recently demonstrated planarized waveguide platform [17] offers a promising solution. This approach, where a low-loss polymer such as benzocyclobutene (BCB) surrounds the laser cavity, would provide the necessary mechanical support to the top metal layer and improve adhesion.

Finally, improvements in the modeling of gain dynamics are essential for understanding the effects of various parameters on device performance. While the Maxwell-Bloch equations are, in principle, exact, their large parameter space makes analysis challenging [18], [19]. Mean-field approaches, although more stable and a good predictive tool, rely on an adiabatic approximation that limits their accuracy, especially for comb bandwidths comparable to or larger than $1/\pi T_1$ where $T_1$ is the gain recovery time. Recently, a generalized mean-field theory has been formulated [20] to address these limitations, which reduce the Maxwell-Bloch equations to a single equation in the non-adiabatic limit, which enables efficient modeling of systems with moderate to large comb bandwidths.

**Concluding Remarks**

The strategy of loss shaping introduces new flexibility in designing active cavity combs, particularly QCL combs. So far, efforts to develop novel comb states have mainly focused on dispersion engineering, and gain engineering has only been done through the creation of broadband heterogeneous designs. While heterogenous QCL designs are crucial for achieving octave-spanning combs, obtaining sufficiently flat gain spectra solely through MBE growth is challenging. Combining multiple stacks inherently introduces gain variations, and even small shifts (a few cm$^{-1}$) can significantly impact performance. Loss and gain shaping offers a way to mitigate these variations and is also compatible with other cavity types.

**Acknowledgements**

D.B. acknowledges the support from AFOSR grant no. FA9550-24-1-0349, ONR grant N00014-21-1-2735, AFOSR grant no. FA9550-20-1-0192, and NSF grant ECCS-2046772; this research is funded in part by the Gordon and Betty Moore Foundation through Grant GBMF11446 to the University of Texas at Austin to support the work of D.B.

## 4.1 Quantum Cascade Laser Feedback Interferometry for Terahertz Imaging

Karl Bertling[1], Jari Torniainen[1], Paul Dean[2] Aleksandar D. Rakić[1]

[1]School of Electrical Engineering and Computer Science, The University of Queensland, St Lucia, QLD 4072, Australia

[2]School of Electronic and Electrical Engineering, University of Leeds, Leeds LS2 9JT, United Kingdom

k.bertling@uq.edu.au, j.torniainen@uq.edu.au, P.Dean@leeds.ac.uk, a.rakic@uq.edu.au

**Status**

Laser Feedback Interferometry (LFI) is a self-mixing technique that utilises the interaction between a laser's emitted and backscattered light to extract target information. When applied to Quantum Cascade Lasers (QCLs) in the terahertz (THz) range, LFI enables high-resolution, non-invasive imaging and sensing [1].

LFI relies on a portion of the emitted beam being reflected back into the laser cavity after interacting with the external cavity (or target), manifest as a small perturbation of terminal voltage across the QCL. By analysing these variations, information about the target's distance, reflectivity, absorption, and surface morphology can be obtained. THz QCLs are ideal sources for LFI due to their high emission power, narrow linewidth, and frequency tunability. The feedback signal directly affects the laser's electro-optical characteristics, allowing real-time imaging without external interferometric setups [2].

Traditional THz detectors, such as bolometers and Schottky diode detectors, often require cryogenic cooling for high detection sensitivity, making them bulky and costly. They also suffer from slow response times and rely on complex external optics, increasing system complexity. In contrast, LFI with QCLs offers a compact, self-contained system that eliminates the need for extra cryogenic cooling while enabling fast, high-sensitivity imaging [3].

Key advantages of LFI with QCLs include its high sensitivity to weak reflections [4], real-time operation, and ability to perform non-contact, label-free imaging. Its compact design simplifies integration as portable systems, making it particularly useful for biomedical diagnostics [5,6], agri-photonics [7], and material and device inspection [8,9]. Additionally, the transparency of materials such as biological tissues, plastics, and semiconductors in the THz range enhances LFI's effectiveness for subsurface imaging. These benefits position LFI with QCLs as a cutting-edge solution for diverse THz imaging applications (see Fig 1.).

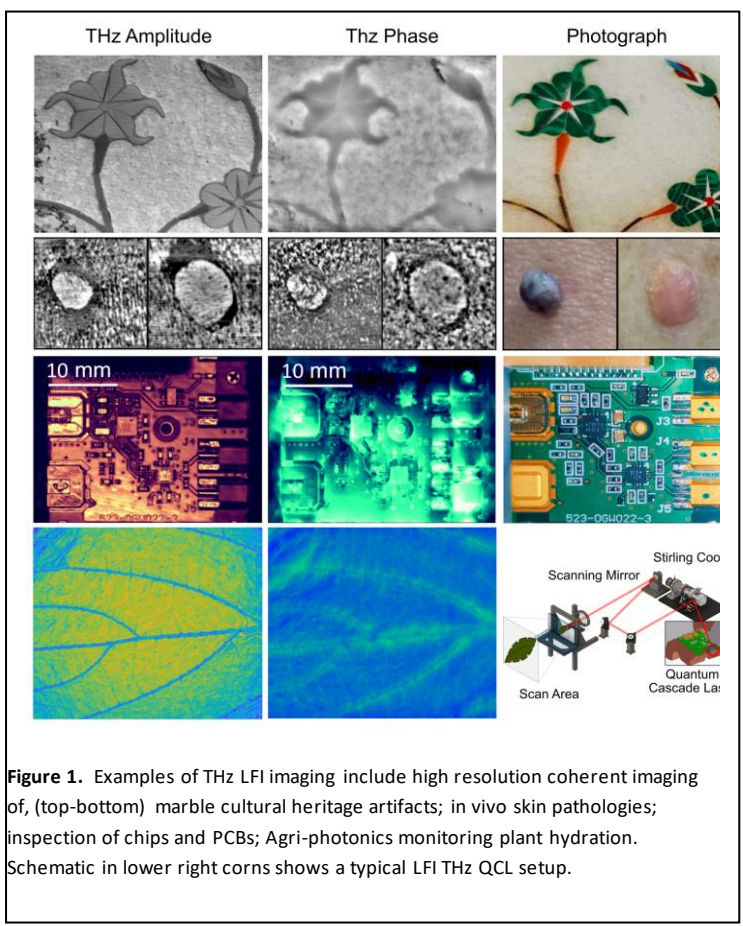

**Figure 1.** Examples of THz LFI imaging include high resolution coherent imaging of, (top-bottom) marble cultural heritage artifacts; in vivo skin pathologies; inspection of chips and PCBs; Agri-photonics monitoring plant hydration. Schematic in lower right corns shows a typical LFI THz QCL setup.

**Current and Future Challenges**
THz QCL LFI imaging is a single-pixel technique that comes with several challenges. Although individual pixel acquisition is fast—limited mainly by the interferometric fringe generation and detection speed [10-12] — the overall imaging process suffers from slow scanning, reducing real-time application capabilities and spatial resolution compared to focal plane arrays. The technique also requires precise optical alignment, making it prone to mechanical instability and noise. The signal-to-noise ratio (SNR) can be low, especially for weak reflections, which necessitates longer integration times to improve image quality. While computational methods like compressive sensing can enhance efficiency, they add complexity and increase processing demands. Additionally, THz QCL LFI imaging faces challenges in uncontrolled environments due to the high absorption of THz radiation by water vapour lines, particularly in humid conditions. Furthermore, QCLs require significant cooling, often operating near cryogenic temperatures, which limits their practical use. Finally, the limited THz bandwidth of QCLs in LFI configurations makes spectroscopic imaging difficult to implement.

**Advances in Science and Technology to Meet Challenges**
Recent advances in pulse-based LFI imaging provide effective solutions to several challenges. The use of sub-microsecond pulses reduces imaging artifacts caused by vibrations and unstable targets, while high-duty cycle operation shortens integration times [13,14]. To address environmental limitations, optical path folding within vacuum enclosures with minimal exposure to humidity helps mitigate THz absorption issues, enabling the deployment of THz LFI instruments outside controlled lab environments. Additionally, the development of non-mechanical scanning techniques and innovative adaptations of existing mechanical methods will expand real-world application scenarios. THz QCLs continue to improve, with operating temperatures approaching room temperature under certain conditions [15]. Even incremental advancements in temperature performance could enhance the competitiveness of these optical sensors by eliminating the need for external cryogenic detectors while preserving high detection sensitivity and low noise. Integrating pulsed QCLs with Stirling cryocoolers offers a compact, efficient solution for real-world THz imaging. Bandwidth limitations can be addressed by using QCL arrays simultaneously operating at different frequencies [16], advanced THz QCL tuning schemes [17] or employing THz frequency combs [18,19].

**Concluding Remarks**
Laser Feedback Interferometry (LFI) with Quantum Cascade Lasers (QCLs) enables high-resolution, non-invasive THz imaging for applications like biomedical diagnostics and security screening. While offering compact, potential real-time imaging, current challenges like slow scanning speeds, sensitivity issues in humid environments, and the need for external cooling systems remain. Recent advancements in pulse-based LFI, non-mechanical scanning, and improved QCL radiation and temperature performance are addressing these limitations, making THz QCL LFI more viable for practical use in foreseeable future.

**Acknowledgements**
*[KB, ADR, & JT Acknowledge funding from Australian Government CTCP (CTCF000040) and NHMRC Development Grant (*2039614*) programs]*

## 4.2 Optical feedback dynamics in quantum cascade lasers

Xiaoqiong Qi[1], Thomas Taimre[2]
1. School of Electrical Engineering and Computer Science, The University of Queensland, Brisbane, QLD 4072, Australia
2. School of Mathematics and Physics, The University of Queensland, Brisbane, QLD 4072, Australia

**Status**

Dynamics are ubiquitous in nature, from the motion of a pendulum to hurricanes. In semiconductor lasers, complex dynamics—including periodic oscillations and coherence collapse—have been observed under external perturbations, such as optical injection, optical feedback, and optoelectronic feedback [1-6]. In 1986, Tkach and Chraplyvy identified five feedback regimes in 1.5 μm distributed feedback lasers [1]. Later studies revealed that the relationship between the relaxation oscillation frequency ($f_{RO}$) and the external cavity resonant frequency ($f_{EC}$) determines whether a laser exhibits chaotic behavior [2]. In the log-cavity regime ($f_{RO} \gg f_{EC}$), five feedback regimes emerge, while in the short-cavity regime ($f_{RO} < f_{EC}$), periodic pulsing and regular pulse packets replace chaos. Compared to diode lasers, quantum cascade lasers (QCLs) have a lower linewidth enhancement factor, reducing the complexity of optical feedback dynamics [7-9]. Nevertheless, similar five-regime feedback behavior has been observed in mid-infrared (MIR) QCLs [10]. Terahertz (THz) QCLs have been considered highly stable under feedback due to their ultrafast carrier lifetimes and lower linewidth enhancement factors than those of MIR QCLs [7]. However, recent studies challenge this assumption, self-pulsations have been experimentally demonstrated and theoretically explored in THz QCLs under optical feedback, with oscillation frequency increasing with increasing feedback strength [9], and titled feedback has also been reported to generate periodic, quasi-periodic oscillations, and square-wave patterns [11]. Optical feedback can also trigger fundamental and harmonic frequency combs in QCLs [12]. These findings highlight that while QCLs are generally robust against feedback, specific conditions can trigger rich dynamics. A deeper understanding of these dynamics is crucial for advancing applications in sensing, imaging, spectroscopy, and optical communication technologies.

**Current and Future Challenges**

Optical feedback dynamics have been extensively studied in various output mode structures, including single-mode MIR-QCLs and THz QCLs [7-11, 13, 14], tunable-mode THz QCLs with coupled cavities [15], multi-mode THz QCLs [16], and optical frequency comb emissions [12]. In multimode THz QCLs, three feedback regions—single-mode, multi-mode, and tunable mode emission have been identified, depend on the gain bandwidth of the laser and feedback strength [16]. Furthermore, single-mode feedback dynamics, such as transient self-pulsations, have been demonstrated for THz imaging under pulse mode [13] (Figure 1b). Compared to conventional THz imaging using laser feedback interferometry (LFI) with laser current sweeping (Figure 1a), self-pulsation-based imaging requires no driving current sweep, making it ideal for the high temperature THz QCLs with limited current modulation range. The self-pulsation dynamics can also modulate self-mixing signals under strong feedback (Figure 1c). For the future applications in optical sensing, imaging, and spectroscopy at MIR and THz frequencies, hyperspectral sensing and imaging are crucial for enabling broadband material characterisation and multi-frequency imaging. However, two main challenges remain: (1) laser operation conditions required to establish a monotonic relationship between the sensing parameter (e.g., target reflectivity) and a measurable physical parameter (e.g., amplitude of the optical feedback dynamics) across multiple frequencies are not well understood; (2) broadband detection remains difficult due to high bandwidth requirements associated with the frequency spectral range (FSR) of QCL cavities, typically on the order of tens of GHz. One solution to reduce the detection bandwidth is using dual-comb spectrometer with slightly different FSR, by detecting the multi-heterodyne beating signal from one of the laser biases, the optical spectrum can be downconverted to lower frequencies at the difference frequency between two laser's FSR while preserving the original optical spectrum [17]. Lasers are inherently nonlinear and exhibiting complex dynamics, including chaos. Investigating these dynamics not only deepens our understanding of fundamental physics but also enables various applications. Chaos have been observed in single-mode MIR QCLs [10], and recently, chaos was experimentally observed in free-running multimode THz QCLs [18]. Understanding these mechanisms could advance their applications while also help to avoid unwanted dynamics in laser systems for communications and precision measurements.

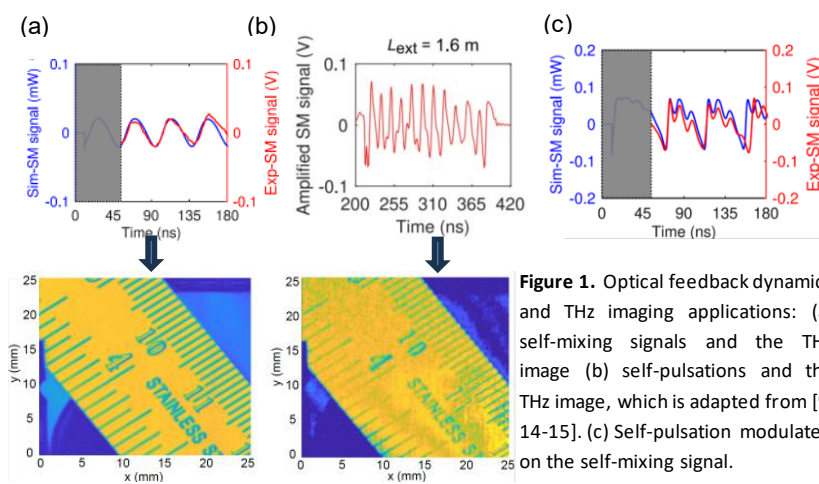

**Figure 1.** Optical feedback dynamics and THz imaging applications: (a) self-mixing signals and the THz image (b) self-pulsations and the THz image, which is adapted from [9, 14-15]. (c) Self-pulsation modulated on the self-mixing signal.

**Advances in Science and Technology to Meet Challenges**

Theoretical models used to investigate the optical feedback dynamics in QCLs are primarily based on reduced rate equations (RREs)[1,3,5,6-9, 11, 13-16], which describes photon-carrier interactions through radiative and nonradiative processes. This model can be used for various mode emission structures and is very suitable for studying time-resolved electron and photon transport dynamics and steady-state analysis. By incorporating the temperature and voltage dependence of the RRE input parameters, the model can also capture the interplay between electrical, thermal, and optical effects [19]. However, RREs lack spatial dependence, although spatial hole burning (SHB) can be approximated through gain modifications. For a more accurate prediction of these effects, Maxwell Bloch equations explicitly consider the polarization of the electric field [12], which describes the interactions between the laser field and the gain medium and includes the effects of Kerr nonlinearities through the optical susceptibility. A recent breakthrough demonstrated pulse-mode lasing of THz QCLs at temperatures up to 250 K [20], advancing portable THz sensing and imaging using thermoelectric coolers.

**Concluding Remarks**

Rich laser dynamics, including periodic oscillations and chaotic oscillations have been observed in MIR- and THz QCLs under optical feedback, with applications in sensing, imaging, and secure communications. Currently most research focuses on optical feedback dynamics in single-mode QCLs for narrow band THz sensing and imaging. However, there is a growing need for broadband techniques which can be achieved by multi-mode, tunable-mode QCLs or optical frequency combs. A deeper understanding of multimode feedback dynamics in QCLs will be the key to future multi-spectral sensing and imaging technologies.

## 4.3 High-speed operation of quantum cascade lasers and their application to conventional and private free-space communications.

Olivier Spitz,[1,*] Frédéric Grillot[2,3,†]

[1]CREOL, College of Optics and Photonics, University of Central Florida, Orlando, FL 32816, USA
[2]LTCI Télécom Paris, Institut Polytechnique de Paris, 19 place Marguerite Perey, Palaiseau, 91120, France
[3]Center for High Technology Materials, University of New-Mexico, 1313 Goddard SE, Albuquerque, NM 87106, USA
[*]Corresponding author: olivier.spitz@ucf.edu
[†]Corresponding author: frederic.grillot@telecom-paris.fr

**Status**

Quantum cascade lasers (QCLs) represent an ideal trade-off for mid-infrared free-space optical communications (FSOCs) as they are less energy consuming than frequency conversion processes [zhou2024] while being able to generate more optical power than interband cascade lasers [didier2023]. Moreover, QCLs can be designed to emit at room temperature across the mid-infrared and long-wave infrared domains. QCLs not only differ from regular interband lasers because they emit in the mid-infrared domain. Their electronic properties are also different and that includes the notable lack of relaxation oscillation frequency [capasso2002]. The latter is known for limiting the modulation speed of all interband semiconductor lasers, including interband cascade lasers [didier2023]. The lack of relaxation oscillation in QCLs was thus considered a major advantage for high-speed communication, in conjunction with the emission in the mid-infrared domain. There are indeed two atmospheric transparency windows [capasso2002, zhou2024] in this optical domain and that converts to a more efficient transmission for a given optical output power, especially when considering ballistic photons. Another key point of the two transparency windows is that there is major band availability (3-5 µm and 8-14 µm) and this means that, in multiplexed schemes, hundreds of parallel QCL spectral lines (from distinct emitters or from a QCL frequency comb [vitiello2022]) can transmit simultaneously, making this yet unlicensed part of the electromagnetic spectrum very appealing.

The initial journey of high-speed FSOCs with QCLs began shortly after the invention of the QCL itself [faist1994]. Data rates of several Gbit/s were already achieved [capasso2002], but this feat was deemed unpractical at that time due to the requirement of liquid nitrogen, leaving the field with only minimal steps forward for more than a decade. The pivotal article that reenergized the field was published in 2017 [pang2017] and marked a major advancement as this was the first demonstration of a room-temperature multi-Gbit/s transmission with a QCL, and that was achieved with direct modulation. Yet, improvements were expected, as this finding was a very short-distance transmission (5 cm) and was heavily relying on pre- and post-processing in order to account for the limited bandwidth of the QCL/detector system.

The next major breakthrough introduced external, amplitude Stark effect modulators operating in the long-wave infrared range [dely2022]. This work was also the first to achieve a transmission rate of 10 Gbit/s, and this without any pre- or post-processing. A similar experiment was subsequently realized, extending the propagation distance to 30 meters thanks to a multi-pass cell and employing signal processing techniques in order to exhibit a maximum data rate of 40 Gbit/s [didier2022]. This paved the way for other amplitude external modulators with bandwidth in the range of 10 GHz [malerba2024], and state-of-the-art result with modulators are presented in Fig. 1. Noteworthy, investigations with direct modulations of QCLs are still ongoing and recently, a 65 Gbit/s transmission has been achieved [dely2024], even if the transmission distance remains limited.

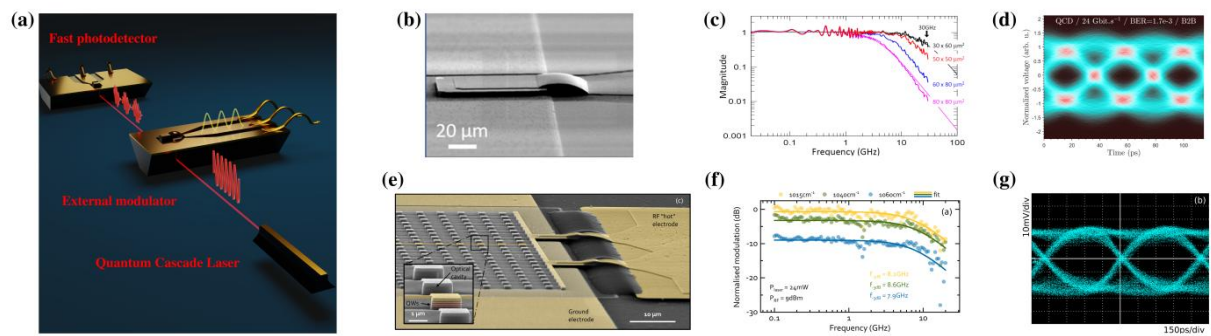

**Figure 1.** Illustration of external modulators for FSOCs with QCLs. (a) Schematic of the three main components of the high-speed transmission. (b) Microscope picture of an amplitude Stark effect modulator. (c) Typical bandwidth of the Stark effect modulators for various sizes of the active area. (d) Eye diagram of a transmission at 24 Gbit/s with a two-level format when the fast detector is a quantum cascade detector (QCD). Adapted from [dely2022] and [didier2022]. (e) Microscope picture of a modulator based on metal-metal optical resonators. (f) Typical bandwidth of the modulator based on metal-metal optical resonators for several wavelengths around 9.6 µm. (g) Eye diagram of a transmission at 1.5 Gbit/s with a two-level format, limited by the bandwidth of the Mercury-Cadmium-Telluride (MCT) detector. Reproduced with permission from [malerba2024].

**Current and Future Challenges**

FSOCs relying on optical carriers are considered less vulnerable to eavesdroppers compared to radiowave transmissions. Nevertheless, interception may still occur and the message to be transmitted is at risk of being read by an illegitimate receiver. Privacy can be implemented within the framework of photonic chaos synchronization. This method requires two twin QCLs with extremely similar properties, one of them being the emitter and the other one the receiver. The emitter is driven into a chaos state, generally with optical feedback from a mirror, and is directly modulated with the small-amplitude message to be transmitted. Recovery of the message after propagation can only occur if the receiver synchronizes with the chaos of the emitter. This method, described in details in Fig. 2, showed a successful private communication up to 8 Mbit/s, with the extra implementation of signal processing [didier2024].

One of the current challenges in both conventional and private communications with QCLs relates to the maximum bandwidth of the transmission. In conventional communications, state-of-the art schemes exhibit maximum bandwidth in the order of 10 GHz. These values are lagging behind the minimum response time exhibited by intersubband detectors at room temperature [lin2023]. On the one hand, this means that the maximum rate of mid-infrared FSOCs will not be limited by the detector capabilities in the near future, but on the other hand, it means that there is space for improvement at the level of the QCL or modulator. If these two devices can match the bandwidth of state-of-the art detectors, then 100 Gbit/s free-space transmissions will be at reach even without using signal processing, which would be a major breakthrough.

Conversely, the maximum rate of private communication is determined by the maximum chaos bandwidth of the emitter. As already emphasized, QCLs do not exhibit relaxation oscillation and that should translate into very large chaos bandwidths, much larger than that of interband lasers. Counter-intuitively, the measured chaos bandwidth of QCLs is, until now, limited to a few tens of MHz [didier2024].

Another challenge lies in the limited propagation distance and indoor environment that characterize almost all the experiments carried out with QCLs so far. A very limited number of decade-old efforts investigated outdoor performance and showed the relevance of the mid-infrared carrier. However, these demonstrations were carried out at low rates [blaser2001,taslakov2008]. With the recent record transmission rates getting close to 100 Gbit/s, the extension to real-field application is highly desired and will be transformative to confirm, for various meteorological conditions, the advantage of mid-infrared wavelengths over competing technologies.

**Advances in Science and Technology to Meet Challenges**

Despite the renewed interest that started 7 years ago, FSOCs with QCLs remains at its early stage and further advances are required before the wide adoption of this technology. Other wavelengths can take advantage of both phase and amplitude external modulators, as well as optical amplifiers. It is likely that the future of QCL communications will rely on those technologies. Recent years have shown progress in amplitude modulators but the development of phase modulators [dely2023] and optical amplifiers [durupt2023] is only at its early stage. Mid-infrared phase modulator technology will be required to unlock the full potential of QCL FSOCs with coherent schemes. Alternatives to the existing technology may arise in the lower part of the mid-infrared domain with SiGe photonic components and with lithium niobate-based modulators. For the latter, experimental demonstrations already reached bandwidths in excess of 10 GHz [lee2024] and this technology can be used up to 6 μm. The main challenge so far remains integration of the lithium niobate modulator on platforms that are compatible with wavelengths > 3 μm.

In order to extend the communication distance to a few hundreds or thousands of meters and thus be compatible with real-field applications, beam shaping is required and can be performed through adaptive optics, which has been used for decades with other wavelengths. Recent investigations have also shown that the effect of turbulence can be cancelled with photonics processors after reception of the FSOC [huang2024]. This concept relies on a photonics mesh designed with several Mach-Zehnder modulators and could decrease the footprint of bulky adaptive optics systems.

The current limited chaos bandwidth in QCLs could be overcome by leveraging array structures. In this configuration, the interaction between the QCLs in the array can give rise to wideband and complex chaotic dynamics. Investigations have focused on numerical simulations so far [grillot2018] but the recent realization of QCL arrays [milbocker2024] can pave the way for more versatile chaos communications, which would not be restricted to the sub 100 Mbit/s range.

Last but not least, QCLs can be envisioned for free-space quantum communications [vitiello2022], yet a new quantum hardware operating at these wavelengths must be invented. For instance, for continuous variable quantum key distribution (CV-QKD), the utilization of coherent states and squeezed states must be properly incorporated into CV QKD protocols.

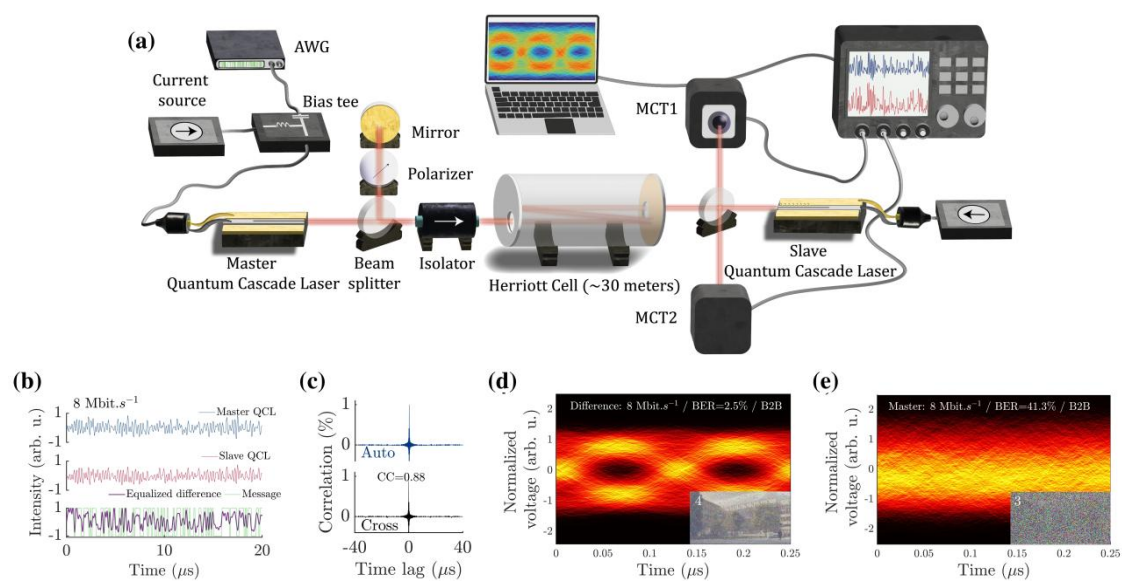

Figure 2. Private communication based on chaos synchronization with two twin QCLs. (a) Experimental setup with the master QCL subject to optical feedback to drive it chaotic and electrically modulated with the message to privately transmitted. The optical waveform with chaos and message is propagated for 30 meters and injected into the slave QCL. The latter synchronizes with the optical chaos only, which means that the difference between the electrical signal of MCT1 and MCT2 allows recovering the concealed message. (b) Intensity time traces of the master QCL (blue), the slave QCL (red) and the equalized difference (purple) with the 8 Mbit/s initial message (green) as a reference. (c) Auto-correlation of the master signal (blue) and cross-correlation between the master signal and the slave signal (black), illustrating the high degree of synchrony. (d) Eye diagram and image recovery (inset) for a legitimate receiver. (e) Eye diagram and image recovery (inset) for an illegitimate receiver. Reproduced with permission from [didier2024].

**Concluding Remarks**

Free-space optical datacom represents a promising alternative to the gradual saturation of channels dedicated to wireless technologies, and to growing bandwidth requirements. There is a clear and strong societal interest in proposing alternatives to counter atmospheric limitations and increase the range of optical telecommunications systems, even in degraded environments. The development of QCLs and associated technologies does have the potential to address secure and high-speed links, which can be further combined with RF communication systems. The design of new components operating at mid-infrared wavelengths, such as modulators, ultra-fast detectors and even optical amplifiers, will play an essential role in the growing development of such FSOC systems. Future performance will also be boosted by the development of coherent communications, photonics integration, wavelength division multiplexing with comb sources and a quantum environment.

**Acknowledgements**

*The Authors acknowledge the financial support of mirSense, the Institut Mines-Télécom, the French National Agency (ANR), the Direction Général de l'Armement (DGA), and the Air Force Office for Scientific Research (AFSOR).*

## 4.4 QCL-based Quartz-Enhanced Photoacoustic Spectroscopy for gas sensing

Marilena Giglio, Angelo Sampaolo, Pietro Patimisco, Vincenzo Spagnolo
PolySense Lab, Dipartimento Interateneo di Fisica, University and Polytechnic of Bari, Via Amendola 173, Bari 70126, Italy
marilena.giglio@poliba.it, angelo.sampaolo@poliba.it, pietro.patimisco@uniba.it, vincenzoluigi.spagnolo@poliba.it

**Status**

Accurate identification and quantification of gaseous compounds plays a crucial role in tackling environmental issues, ensuring industrial safety, and advancing medical diagnostics [1]. Laser absorption-based (LAS) techniques have gained attention for their high selectivity, sensitivity, cost-effectiveness, and portability, enabling real-time, in-field measurements [2] and advancement of quantum cascade lasers (QCLs) enabled the detection of numerous substances at very low concentration levels [3]. Photoacoustic spectroscopy is an indirect LAS technique, using a microphone to detect the acoustic waves generated in a gas sample by the non-radiative energy relaxation of the molecules excited by a modulated laser source. In 2002 the replacement of microphones by 32 kHz quartz tuning forks (QTFs) with high quality factors improved the sensitivities of the newborn quartz-enhanced photoacoustic spectroscopy (QEPAS) technique [4]. Recent custom QTFs with lower resonance frequencies and optimized dimensions further enhanced detection of slow-relaxing molecules, multi-gas species, and long-wavelength excitation gas species [5].

The compactness, high power, narrow linewidth, and tunability of QCLs have significantly boosted QEPAS performance in the mid-infrared and terahertz ranges, achieving excellent noise equivalent concentrations (see Fig. 1) for gases critical to environmental, industrial, and medical applications.

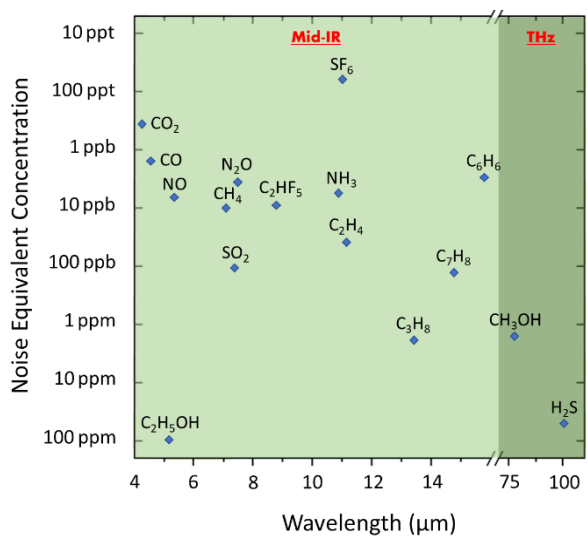

**Figure 1.** Best Noise Equivalent Concentrations obtained with QCLs-based QEPAS sensors in the mid-IR and THz spectral ranges.

**Current and Future Challenges**

Gas sensing often requires detecting multiple species with broad absorption features. QEPAS, with its wavelength-independent QTF responsivity, is well-suited but requires appropriate QCL sources and detection strategies. External cavity QCLs, though effective for broadband sensing, face limitations in size and stability [6]. A recent innovation integrates an array of pulsed distributed-feedback (DFB) QCLs on a single semiconductor chip, offering compactness, broad tunability, and stability [7]. $N_2O$

and $CH_4$ mixtures were analysed over a 150 cm$^{-1}$ spectral range with overlapping absorption bands but required lengthy measurement times for environmental monitoring [8]. For broader or widely spaced features, multi-QCL sources are essential [9]. A recent QEPAS sensor employs a three-wavelength laser module with distinct QCL chips and a collimated beam output, targeting $NO_2$, $SO_2$, and $NH_3$ for environmental monitoring [10]. The provided examples rely on sequential device switching but several applications require the simultaneous detection of different molecules [11], [12], [13]. A dual-laser strategy addressed this by exciting QTF fundamental and overtone modes simultaneously from opposite sides of the resonator [14].

Selectivity remains a challenge when analyzing gas mixtures with overlapping mid-infrared spectral features, such as volatile organic compounds used in medical diagnostics and environmental monitoring. Recently, QCLs emitting at 12–15 µm enabled QEPAS sensors to detect BTEX and lighter hydrocarbons due to distinct spectral features in this range [15], [16]. However, low laser power limits detection sensitivity, making it inadequate for diagnostics.

**Advances in Science and Technology to Meet Challenges**

Real-world gas sensing applications require cheap, sensitive, compact, portable detectors, capable of multi-gas, possibly simultaneous, measurements. QCL-based QEPAS shows promise in meeting these needs, but further advancements in laser sources and QTF development are essential.

QCLs are the most expensive component in QEPAS sensors. Developing slot waveguide QCL arrays, with simpler and robust fabrication, could reduce costs compared to traditional DFB arrays [17].

Enhancing QCL performance at long wavelengths requires addressing low optical gain, carrier leakage, and waveguide losses. Strategies could include improving active region and waveguide designs, using InP-based QCLs, applying diagonal or three-phonon-resonance designs, introducing strain to selectively layers, and buried heterostructure by InP regrowth [18], [19].

For greater ruggedness, dual-gas detection could be integrated onto a single chip, by growing the two different QCLs with a spacing matching the distance between the two QTF antinodes. Furthermore, integrated photonics offers potential for ultra-compact QEPAS sensors: the demonstration of lithium niobate tuning forks with comparable performances with respect to standard QTFs pave the way for photonic gas sensing on LiN platforms [20].

**Concluding Remarks**

The identification and quantification of gaseous compounds are crucial for addressing environmental challenges, ensuring industrial safety, and advancing medical diagnostics. QCLs-based QEPAS has demonstrated exceptional sensitivity, selectivity, and potential for portable, real-time gas sensing. However, challenges such as limited power at long wavelengths, and high laser source costs remain. Developments in slot waveguide QCL arrays, advanced active region designs, and integrated photonics promise to address these limitations, making QEPAS a more versatile, cost-effective, and high-performance solution for gas sensing.

**Acknowledgements**

Authors acknowledge funding from the European Union's Horizon 2020 research and innovation program under grant agreement No. 101016956 PASSEPARTOUT, in the context of the Photonics Public Private Partnership, project MUR – Dipartimenti di Eccellenza 2023–2027 – Quantum Sensing and Modelling for One-Health (QuaSiModO), PNRR MUR project PE0000023-NQSTI and from PNRR MUR project PE0000021-NEST.

## 4.5 THz quantum cascade laser for near-field imaging

Eva A. A. Pogna[1] Istituto di Fotonica e Nanotecnologie, Consiglio Nazionale delle Ricerche (CNR-IFN), P.zza L. da Vinci 32, 20133 Milano, Italy
[eva.pogna@cnr.it]

**Status**

Terahertz quantum cascade lasers (THz-QCLs) have significantly contributed to the recent development of far-infrared near-field imaging, enabling unprecedented nanoscale exploration of light-matter interactions [1-3] and advanced characterization of materials and devices in the THz range (~2-5 THz) [4-7]. Their success stems from their ability to emit intense (mW-level) continuous wave (CW) THz radiation with exceptional spectral purity (~100 Hz linewidth) and to perform phase-sensitive detection via self-mixing interferometry [8-10]. This homodyne detection scheme enables amplitude and phase retrieval with high signal-to-noise ratios, and efficient suppression far-field background, leveraging the high sensitivity and picosecond response time of THz-QCLs.

By coupling the THz-QCLs with sub-wavelength apertures [11] or scattering near-field probes [6], spatial resolutions below the diffraction limit have been achieved, with scattering probes offering the highest precision, down to tens of nanometres [1-10].

THz-QCL-based nanoimaging has enabled mapping of dielectric properties and charge carrier distributions in emerging low-dimensional quantum materials and nanodevices, such as nano-transistors [7], semiconducting nanowires [7], 2D crystals [3,6] (*e.g.* black-phosphorus), topological insulators [1-2,10], and metasurfaces [12]. Additionally, THz-QCL-based scanning near-field optical microscopes (SNOMs) have been utilized to map in-plane propagation of THz polaritons in layered crystals [1-3] and to perform photocurrent nanoscopy studies of THz photoconductivity [7].

Since the first demonstration in 2016 [4], three types of THz-QCLs have been employed in near-field microscopes: (i) single-mode Fabry-Perot THz-QCLs [6-12], (ii) multimode surface-emitting random lasers [13-14], and (iii) THz-QCL frequency combs for hyperspectral nanoimaging [15].

The nanoimaging and spectroscopic capabilities unlocked by THz-QCLs hold immense interdisciplinary potential across physical, chemical and biological sciences and can guide the design of nano-optoelectronic and -photonic devices for next-generation sensing and communication technologies, cementing the role of THz-QCL as key enablers of progress in THz science and technology.

**Current and Future Challenges**

Applying THz-QCLs in near-field imaging faces several technological challenges including:

- **Spectral coverage**, crucial to broaden the applications, targeting material vibrational and electronic resonances across the full THz-QCL emission range (~1.2–5.4 THz). Hyperspectral nanoimaging with THz-QCL frequency combs has achieved a record 680 GHz band at 3.1 THz but detecting only few laser modes [16]. Challenges include maintaining robust comb operation and phase coherence under temperature fluctuations, current hysteresis, and feedback injection [15].
- **Spatial resolution,** of 10s of nanometres over broad frequency range is challenging, due to the frequency dependence of near-field probes' scattering efficiency [17]. Improving requires optimizing tip shape and materials and the post-processing methods for background suppression.
- **Scanning speed,** is practically constrained by THz detector sensitivity and the need to sample the self-mixing fringes for amplitude and phase retrieval. Synthetic holography approach [18] shows promise for faster hyperspectral maintaining spatial and spectral resolution.

- **Cryogenic environment.** Precise control of driving current and heat sink temperature is crucial to maintain frequency and phase stability during THz-QCL CW emission and detection. The dependency on cryogenic liquids poses logistical and economic barriers.

**Advances in Science and Technology to Meet Challenges**

Advancing the application of THz-QCLs in near-field imaging demands innovations in laser technology. The capabilities of THz-QCLs in near-field imaging appear still underexplored. The recent advances in THz-QCL design towards CW operation at higher temperatures with thermoelectric or mechanical coolers could improve the technology's accessibility and sustainability. Improvements in the THz-QCL beam quality, and in the design of near-field probe, are expected to boost signal strength, spatial resolution and spectral coverage. Theoretical and experimental investigations of the effect of optical feedback on THz-QCL frequency comb, under varied operational and feedback conditions [19], could help optimizing the acquisition strategy, enhance performances. Additionally, the GHz beat notes of frequency combs shows sensitivity to optical feedback [15] that could be harnessed for translating the near-field THz scattering information to radio frequencies, accessible to electronics. While, harnessing the phase relationship between the comb modes, e.g. via dual-comb spectroscopy, could unlock new capabilities for advanced nanoimaging. The spectral coverage and portfolio of materials of application, could be expanded by the development of alternatives to the III-V semiconductors commonly used to fabricate QCL, emitting within the 6-12 THz band and supporting higher operating temperature.

THz-QCL based nanoscopy anticipates intriguing application perspectives including: (i) the study of THz polaritons in emerging media with reduced propagation attenuation for THz field sub-diffractional confinement and enhancement; (ii) the investigation of field concentration in individual nanoresonator and metasurfaces for boosting nonlinear light-matter interaction; (iii) the characterization of THz photodetection mechanisms in novel conductive materials; (iv) the advanced testing of nanodevices, *e.g.* mapping photocurrent pathways, charge carrier distribution, inhomogeneiteis in the local structure and THz vibrational dynamics. THz nanoimaging also promises advancements in biological applications, such as tissue analysis and cell-scale imaging, leveraging the high sensitivity to water and the THz molecular vibrational absorptions.

**Concluding Remarks**

THz-QCLs offer unique capabilities for THz near-field imaging, complementing alternative sources like photoconductive antennas, gas laser, free-electron lasers and synchrotron sources [15]. Addressing challenges, such as eliminating the need for cryogenic liquids, stabilizing frequency combs operation, and optimizing scanning and referencing methods —could enhance their performances (spatial and spectral resolution) reducing the operational complexity, the enviromental impact and enable commercial integration. The foreseen research opportunities in near-field imaging position THz-QCLs as a cornerstone technology for material science, nanotechnology and nanophotonics, biomedical nanoimaging enabling significant progress in the analysis of complex nanoscale systems.

**Acknowledgements**

*E.A.A.P acknowledges support from the European Union through the project "Fondo PRIN 2022" – TeRahertz Polaritons unveiled by NEar-field nanoscopy (TRAPNE), contract no. 2022L3KF4S – CUP: B53D2300429 0006 and through the ERC Starting Grant TREAT (GA no.101162914).*

## 4.6 Advances in Near Field Optical Microscopy Based on Infrared Quantum Cascade Lasers

Xiao Guo[1], Rainer Hillenbrand[2,3], Mengkun Liu[4,5], Michael Brünig[1], Adrian Cernescu[6], Alexander A. Govyadinov[6], Karl Bertling[1]

[1]School of Electrical Engineering and Computer Science, The University of Queensland, St Lucia, QLD 4072, Australia
[2]CIC nanoGUNE BRTA and EHU/UPV, Donostia-San Sebastián, Spain
[3]Ikerbasque, Basque Foundation for Science, Bilbao, Spain
[4]Department of Physics and Astronomy, Stony Brook University, Stony Brook, NY, USA
[5]National Synchrotron Light Source II, Brookhaven National Laboratory, Upton, NY, USA
[6]Attocube systems AG (neaspec), Eglfinger Weg 2, 85540 Haar (München), Germany
xiao.guo@uq.edu.au, r.hillenbrand@nanogune.eu, mengkun.liu@stonybrook.edu, m.bruenig@uq.edu.au, adrian.cernescu@attocube.com, alexander.govyadinov@attocube.com, k.bertling@uq.edu.au

**Status**

Scanning near-field optical microscope is a phase-sensitive technique that exploits optical near field, *i.e.*, non-propagating inhomogeneous evanescent fields, to probe local material properties. In scattering-type SNOM (s-SNOM), the spatial resolution is determined by the tip radius: incident radiation is concentrated at a tip apex to form a nanoscale hotspot, which interacts with samples. This tip-sample interaction depends on local material properties and modifies tip scattered field. By recording tip scattered field, s-SNOM allows mapping local optical parameters (*e.g.*, permittivity and conductivity), from which chemical compositions and structural properties can be inferred [1], [2].

The choice of s-SNOM light source is critical. Quantum Cascade Lasers (QCLs), particularly InP-based mid-infrared (mid-IR, 400 to 4000 $cm^{-1}$ or 50 to 500 meV) devices, provide coherent radiation with high output power (> 12 W for continuous wave operation), narrow linewidth (< 1 $cm^{-1}$), and high-speed tuning (> $3 \cdot 10^4$ $cm^{-1}$/sec, Daylight Solutions MIRcat) [3], [4]. In contrast to terahertz (4 to 400 $cm^{-1}$ or 0.5 to 50 meV) QCL detection, which often relies on swept-frequency delayed self-homodyning [5], mid-IR QCLs are naturally compatible with local oscillator phase-modulation schemes such as pseudo-heterodyne detection [2], enabling background-free near-field optical measurements.

The mid-IR range covered by QCLs overlaps with several fundamental excitations: the molecular fingerprint region [6], plasma frequency of doped materials, Reststrahlen bands of polar dielectrics [7], and polariton gaps across diverse physical systems [8]. Consequently, QCL-based s-SNOM stands out to visualise the nanoscale distribution of distinct functional groups in individual macromolecules [9], map Drude response across conductive-insulating domains in 2D polymers [10] and visualise the tuneability of surface polariton wavefronts in anisotropic crystals [11, 12] and quantum materials, *e.g.*, magnetic-field control of polariton topological transitions beyond the limit of electrostatic gating [13].

**Current and Future Challenges**

One main challenge for QCL-based s-SNOM is the restricted spectral tuning range: a single QCL chip spans < 300 $cm^{-1}$ in the mid-IR region [3] and commercial modules integrate up to 4 chips (*e.g.*, Daylight Solutions MIRcat). Frequencies above 2500 $cm^{-1}$, essential for probing N-H and O-H stretches in biological systems [14], and below 750 $cm^{-1}$, relevant for protein-water dynamics [15] confinement-driven anomaly (*e.g.*, ultrathin thickness-induced polariton mode splitting [16]), remain largely inaccessible. Another challenge is searching unique low-photon-energy quantum phenomena under extreme environments such as liquid, cryogenic temperatures, and strong magnetic fields. Most applications remain limited to steady-state measurements of solid samples at room temperature, owing to reduced signal-to-noise ratios or experimental complexity. In addition, interpreting QCL-based s-SNOM signals on complex samples remains a non-trivial challenge. The complex tip-sample interaction complicates quantitative analysis, especially in heterogeneous or topographically intricate systems. Key open questions include how to: (1) leverage artificial intelligence (AI) to quantify tip-sample interactions, (2) use AI-assisted digital twins for mesh-free simulations to bridge the scale mismatch between probe tips and optical wavelengths, and (3) automate and standardise s-SNOM calibration and data analysis with benchmarks reliable across labs worldwide?

**Advances in Science and Technology to Meet Challenges**

A promising path forward lies in the development of broadband mid-IR QCL emitters. Recent work has demonstrated monolithic photonic integrated circuits that multiplex dissimilar QCL active regions on a single InP wafer. By on-chip multiplexing, a single device can combine emission spanning > 800 cm$^{-1}$ spectral range [4]. Another strategy is integrating s-SNOM with optical parametric oscillators (OPOs), which features a broad tuning range (555 to 7140 cm$^{-1}$) with moderate linewidth (*ca.* 4 cm$^{-1}$), *e.g.*, a recent study separated carrier and mobility contributions to plasma frequency spectral shifts in a 2D electron gas at 8 K [17], while QCL can monitor ultrafine control of carrier density and the intersubband transition in cascaded quantum wells, for future on-chip quantum sensing applications. A recent paradigm shift demonstration in liquid IR s-SNOM shows how to track UV/blue light-induced structural switching of lipid nanocarriers in liquids, capturing single-liposome isomerisation dynamics [18], which can extend to study other systems by broadband QCL-based s-SNOM, *e.g.*, mapping local hydrogen bonding strength distribution, measuring bond vibrational lifetime (ultrafast fs mid-IR pump probe), tracking proton transfer pathways, or electronic state excitations (visible pump). Future static-state signal enhancement can be pursued via nanostructured superlens, impedance-matching layers, or active resonators [19] integrated with QCL gain media, amplifying weak evanescent fields.

Additionally, data-driven s-SNOM analysis, while still in its infancy, is emerging. Recently, CycleGAN has been used to denoise low-integration-time QCL-based s-SNOM images (3.3 ms/pixel) (cell slice) into high-fidelity images comparable to threefold longer acquisitions [20]. The next frontier lies in physics-informed neural networks to embed Maxwell equations into learnable architectures (*e.g.*, Kolmogorov–Arnold Networks), yielding interpretable digital twins of electromagnetic fields in the full space–time domain to capture spatial inhomogeneity with a mesh-free collocation method.

**Concluding Remarks**

QCL-based s-SNOM is evolving into a powerful platform that delivers label-free insights into nanoscale science. Its compatibility with liquids, cryogenic states, high magnetic fields, and ultrafast temporal resolution uniquely places it to explore exotic light-matter interactions. By coupling with AI-assisted, high-throughput approaches, next-generation s-SNOM will redefine how we interrogate quantum materials, nanophotonic structures, and biological systems at the nanoscale.

**Acknowledgements**

*The authors acknowledge support from Australian Government CTCP (CTCF000040) and NHMRC Development Grant (2039614) programs. M.K.Liu acknowledges Gordon and Betty Moore Foundation DOI: 10.37807/gbmf12258 for supporting the development of cryogenic IR SNOMs. R.H. acknowledges support from Grant CEX2020-001038-M funded by MICIU/AEI /10.13039/501100011033 and Grant PID2021-123949GB-I00 (NANOSPEC) funded by MICIU/AEI/10.13039/501100011033 and by ERDF/EU.*

## 4.7 Present and Future Challenges for Mid-/Far-Infrared Quantum Light Generation via Quantum Cascade Lasers

Tecla Gabbrielli[a,b], Jacopo Pelini[a,b,c], Irene La Penna[a,b], Alessia Sorgi[a,b], Paolo De Natale[a,b], Davide Mazzotti[a,b], Iacopo Galli[a,b], Luigi Consolino[a,b], Francesco Cappelli[a,b], and Simone Borri[a,b]

[a]CNR-INO - Istituto Nazionale di Ottica, Via Carrara, 1 - 50019 Sesto Fiorentino FI, Italy
[b]LENS - European Laboratory for Non-Linear Spectroscopy, Via Carrara, 1 - 50019 Sesto Fiorentino FI, Italy
[c]University of Naples Federico II, Corso Umberto I, 40 - 80138 Napoli, Italy
*email address of all the authors*[1]

**Status**

Since their first demonstration 30 years ago, quantum cascade lasers (QCLs) have gained a central role in driving mid- and far-infrared (MIR and FIR) research towards precision applications, with key advantages like compactness, up to W-level output power in CW operation at room temperature, narrow intrinsic linewidth [1], and direct frequency comb emission [2]. Thanks to their peculiar characteristics, QCLs are nowadays the most widely used mid-IR source for a variety of applications, including trace-gas detection [3], free-space communication [4,5], and imaging [6].

By pushing the research towards new frontiers of sensitivity and precision, and following the quantum revolution that has already invested the research at shorter (e.g., telecom) wavelengths, we are nowadays witnessing an increasing curiosity and significant efforts in the development of suitable tools to generate and detect nonclassical states of light in the MIR and FIR regions.

In this scenario, QCLs emerge as compact chip-scale candidates, suitable for integration into photonic circuits. The high third-order nonlinearity of their active medium, responsible for their comb emission [2,7,8], triggers four-wave-mixing (FWM) multiseed processes. In other comb-based photonic platforms, FWM is correlated with the generation of squeezed and entangled states of light [9,10].

Triggered by these potentialities, in recent years several efforts have been made to equip the MIR with adequate detection tools for characterizing non-classical states of light. A significant step forward was represented by the first demonstration of a MIR balanced detection system based on state-of-the-art MCT photodiodes with 40% quantum efficiency and able to measure sub-shot-noise signals in the 4-5 µm spectral region with a clearance of up to 9 dB over a bandwidth of 100 MHz [11]. The same balanced detection setup has been used in combination with a diffraction grating (Fig.1) to prove, for the first time, the presence of classical intensity correlations in twin modes generated by FWM starting from a strong pump (three-mode Harmonic Comb, HC) generated by MIR QCLs operating in the 4-5 µm region (Fig.1) as described in ref. [12].

Moreover, these works are actively promoting the theoretical [13, 14] and experimental investigation of new-generation devices [15] targeting the generation and manipulation of nonclassical MIR states of light.

---

[1] tecla.gabbrielli@ino.cnr.it, pelini@lens.unifi.it, lapenna@lens.unifi.it, alessia.sorgi@ino.cnr.it, paolo.denatale@ino.cnr.it, davide.mazzotti@ino.cnr.it, iacopo.galli@ino.cnr.it, luigi.consolino@ino.cnr.it, francesco.cappelli@ino.cnr.it, simone.borri@ino.cnr.it

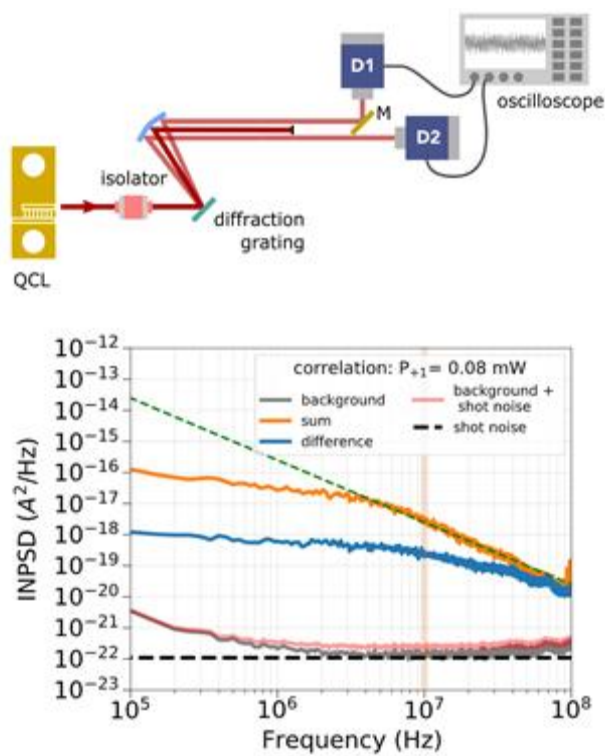

Figure 1. On top: sketch of the balanced detection scheme used to measure correlations. A diffraction grating splits the three modes of a harmonic comb emitted by a QCL, and two MIR photodiodes detect the intensity noise of each side mode. By comparing the sum and the difference of the obtained signals, it is possible to establish the presence of correlation. On bottom: intensity noise power spectral density of the sum (orange trace) and the difference (blue trace) of the two sidebands' measured signals compared to the background noise (grey trace), expected shot-noise level (black dashed lines), and the sum of these two latter contributions (red trace). Figures adapted from [13]. The green line shows the trend of the noise decay, which is expected to reach the relative shot noise around 1 GHz.

**Current and Future Challenges**

Despite the increasing interest, developing quantum tools in the MIR and FIR leads to unavoidable challenges. As discussed in ref. [12], the MIR still lacks the technological maturity that characterizes shorter wavelengths, where compact setups for generating quantum light states are already available [9,16]. At present, one of the limiting factors preventing the detection of FWM-generated nonclassical correlations in QCLs is detectors' performance. Among the available MIR detectors, the technological challenge comes from the possibility of combining, in the same device, a high-enough quantum efficiency, a high bandwidth, and good dynamical range and clearance [11]. The results reported in ref. [12] suggested that correlation measurements would require detectors with high bandwidth (around a few GHz) to avoid the laser residual noise overwhelming the shot-noise at low Fourier frequencies (Fig.1, green trend). However, the low saturation level of fast detectors, due to their smaller area, generally compresses the available dynamical range, preventing the detection of sub-shot-noise signals.

The technological challenges of this spectral region also affect the laser sources. Firstly, the possibility of detecting sub-shot noise signals crucially depends on the possibility of having a laser intensity noise dominated by Poissonian statistics, with a limited amount of excess noise that needs to be ruled out in differential measurements. For QCLs, theory predicts a flat intensity noise spectrum up to Fourier frequencies of about 1 GHz for low/moderate emitted power (1-100 mW) [17]. However, additional technical factors like driving-source noise and high-frequency noise of QCLs in combination with the $\alpha$-factor have been indicated as possibly responsible for the presence of excess noise in the relative intensity noise (RIN) at low Fourier frequencies [18]. Moreover, the survival of nonclassicalities in a complex nonlinear system such as a laser device is not guaranteed, as the active gain can be responsible for their degradation. The results shown in ref. [12] seem promising in terms of correlation survival; however, the detected signal suffers from a significant residual noise in the tested bandwidth (approximately 100 MHz), preventing it from reaching even the related shot-noise level (Fig. 1).

These challenges are even more critical in the FIR (1-10 THz) region, where the technology is nowadays not mature enough to detect nonclassical states of light. Even though a huge work has been carried out from the sources point of view, resulting in an engineered QCL HC generation [19], the bottleneck is represented by the detectors availability. In fact, the most sensitive and fast commercially available FIR detectors, such as hot electron bolometers, are not suitable for this task. The typical values reported by the manufacturers of dynamical ranges (in the order of 0.1 µW), combined with a NEP of about $7 \cdot 10^{-14}$ W/Hz$^{-1/2}$, do not allow sub-shot-noise-level detection.

**Advances in Science and Technology to Meet Challenges**

One of the most significant efforts for developing adequate quantum tools in the MIR was represented by the Qombs project [GA 820419, doi: 10.3030/820419], funded under the European Quantum Flagship initiative [20]. Starting from quantum simulations performed with ultracold atoms, the project was aimed at designing and engineering a new generation of QCL frequency combs. This project and related research activities [11,12] allowed for the development of quantum tools in the MIR and offered insights on how to overcome the technological challenges related to these wavelengths.

As reported in ref. [12], correlation measurements in QCLs are affected by a significant residual noise from the sources, preventing them from reaching even the shot-noise level. To solve this problem, different paths need to be travelled in parallel. New detectors should be designed *ad hoc* with optimized features like bandwidth, saturation level, quantum efficiency, common-mode rejection ratio capability, and a good dynamical range. This is required to reach a good clearance, thus matching the high standards required by quantum measurements with bright states. Moreover, comparing the intensity noise properties of different MIR state-of-the-art devices [15] can help in identifying the most promising low-intensity-noise sources, which can be exploited for the generation of MIR shot-noise-limited or even sub-shot-noise radiation. For this purpose, a deep investigation of the dynamic processes in QCLs is needed to understand the key factors that affect their intensity noise, paving the way for the fabrication of a new generation of quantum-optimized sources.

Finally, all the mentioned works suggest a path to bring these technologies to FIR frequencies, where the technological gap is even more pronounced. Here, the challenges of transferring quantum-technology platforms to the FIR frequency range is the aim of the QATACOMB project, funded by the QuantERA program [GA 101017733, doi: 10.3030/101017733]. The goal is to develop a complete system for detecting and characterizing nonclassical light states, leveraging THz QCL-HCs as nonlinear sources and novel graphene-based quantum sensors. To date, a semiclassical simulation framework for modeling the spatiotemporal dynamics in active photonic devices, such as QCLs, has been proposed [14] and applied to simulate HCs in THz QCLs. The obtained results well reproduce the MIR HC intensity noise trends as in ref. [12]. This first result shows great potential for the theoretical investigation of intermodal intensity correlations, in view of revealing nonclassicalities.

**Concluding Remarks**

The worldwide quantum revolution has ultimately reached longer wavelengths, involving both the MIR and FIR spectral regions. QCLs have gained attention as promising chip-scale coherent sources that could emit *per se* nonclassical states of light due to the high third-order nonlinearity of their active medium. To address the lack of quantum tools in these spectral ranges, many efforts have been undertaken, especially in the MIR, to provide appropriate detection tools and sources. This has led to the realization of the first balanced detection scheme capable of measuring sub-shot-noise-level signals, the detection of FWM-triggered correlation in HCs emitted by QCLs, and the intensity noise characterization of diverse MIR state-of-the-art devices. A more recent advancement is represented by the demonstration of few-photon coherent detection from a cw MIR QCL using a balanced detector based on commercial MIR photodiodes [20], with interesting perspectives towards room-temperature MIR single-photon detection. These works have started to populate the MIR region with technologies suitable for the detection and characterization of quantum sources, and give insightful hints toward the realization of the first cascade-based laser capable of directly generating nonclassical states of light.

**Acknowledgements**

The authors acknowledge financial support by the European Union's NextGenerationEU Programme [doi: 10.2761/808559] with the I-PHOQS Infrastructure [IR0000016, ID D2B8D520, CUP B53C22001750006] "Integrated infrastructure initiative in PHOtonic and Quantum Sciences". The authors also acknowledge financial support from the Horizon Europe MUQUABIS Project [GA 101070546, doi: 10.3030/101070546] "Multiscale quantum bio-imaging and spectroscopy", from the Horizon 2020 Laserlab-Europe Project [GA 871124, doi: 10.3030/871124], from the European Union's QuantERA II QATACOMB Project [GA 101017733, doi: 10.3030/101017733] "Quantum correlations in terahertz QCL combs", from the Italian Ministero dell'Università e della Ricerca (project PRIN-2022KH2KMT QUAQK), from ASI and CNR under the Joint Project "Laboratori congiunti ASI-CNR nel settore delle Quantum Technologies (QASINO)" (Accordo Attuativo n. 2023-47-HH.0) and from the EURAMET 23FUN04 COMOMET Project [doi: 10.13039/100019599].